\documentclass{article}
\usepackage[utf8]{inputenc}
\usepackage{physics}
\usepackage{amsthm}
\usepackage{amsmath}
\numberwithin{equation}{section}
\usepackage{amssymb}
\usepackage[margin=1.2cm]{geometry}
\usepackage{tcolorbox}
\usepackage{simpler-wick}
\usepackage{xcolor}
\usepackage{slashed}
\usepackage{graphics}
\newif\ifnatbibsort\natbibsorttrue
\relax
\ifnatbibsort\RequirePackage[numbers,sort&compress]{natbib}\else\RequirePackage[numbers,compress]{natbib}\fi
\RequirePackage[colorlinks=true
  ,urlcolor=blue
  ,anchorcolor=blue
  ,citecolor=blue
  ,filecolor=blue
  ,linkcolor=blue
  ,menucolor=blue
  ,linktocpage=true
  ,pdfproducer=medialab
  ,pdfa=true
]{hyperref}

\newcommand{\om}{\omega}
\newcommand{\oemga}{\omega}

\date{}
\usepackage{authblk}

\newsavebox\affbox
\author{Dhruva K.S\footnote{\href{k.s.dhruva@students.iiserpune.ac.in}{\texttt{k.s.dhruva@students.iiserpune.ac.in}}}$^1$,\,
\,Deep Mazumdar\footnote{\href{deepkamal.mazumdar@students.iiserpune.ac.in}{\texttt{deepkamal.mazumdar@students.iiserpune.ac.in}}}$^1$, and Shivang Yadav\footnote{\href{shivang.yadav@students.iiserpune.ac.in}{\texttt{shivang.yadav@students.iiserpune.ac.in}}} }
\affil[1]{ \textit{Indian Institute of Science Education and Research,} \par\textit{
Dr. Homi Bhabha Road, Pashan, Pune, India}
} 
\date{}    
\begin{document}
\title{\textbf{
\textit{n}-point functions in Conformal Quantum Mechanics:\\A Momentum Space Odyssey
}}

\maketitle
\begin{abstract}
    In this paper, we study the implications of conformal invariance in momentum space for correlation functions in quantum mechanics. We find that three point functions of arbitrary operators can be written in terms of the $_2 F_1$ hypergeometric function. We then show that generic four-point functions can be expressed in terms of Appell's generalized hypergeometric function $F_2$ with one undetermined parameter that plays the role of the conformal cross ratio in momentum space. We also construct momentum space conformal partial waves, which we compare with the Appell $F_2$ representation. We test our expressions against free theory and DFF model correlators, finding an exact agreement. We then analyze five, six, and all higher point functions. We find, quite remarkably, that $n$-point functions can be expressed in terms of the Lauricella generalized hypergeometric function, $E_A$, with $n-3$ undetermined parameters, which is in one-to-one correspondence with the number of conformal cross ratios. This analysis provides the first instance of a closed form for generic momentum space conformal correlators in contrast to the situation in higher dimensions. Further, we show that the existence of multiple solutions to the momentum space can be attributed to the Fourier transforms of the various possible time orderings. Finally, we extend our analysis to theories with $\mathcal{N}=1,2$ supersymmetry, where we find that the constraints due to the superconformal ward identities are identical to identities involving hypergeometric functions. 
\end{abstract}
\newpage
\tableofcontents
\newpage
\section{Introduction}
The conformal bootstrap program is a highly successful endeavor to obtain nonperturbative results in quantum field theory. In two dimensions, this program was initiated in the seminal paper by Belavin, Polyakov and Zamolodchikov in 1984 \cite{Belavin:1984vu}. In dimensions greater than two, there has also been significant progress since the initiating work \cite{Rattazzi:2008pe} in 2008, see \cite{Hartman:2022zik,Poland:2022qrs} for a review of recent developments. In contrast to conformal field theory (CFT) in position space and mellin space, momentum space CFT is less understood. In the hope of furthering our understanding of momentum space CFT, we focus on the one-dimensional case, i.e., on conformal quantum mechanics (CQM) in this paper. Compared to its higher dimensional counterparts, it provides us a path with fewer technical hurdles to obtain exact analytical results such as closed form expressions of four and higher point point functions. Historically, the study of conformal quantum mechanics dates back to 1976, with the work of de Alfaro, Fubini and Furlan (DFF)\cite{deAlfaro:1976vlx} who analyzed a particular quantum mechanical model (A particle moving in a inverse square potential) that possesses conformal invariance. Almost five decades after this paper, there has been a great deal of development with applications ranging from the connection of CQM to M2 branes, black holes and even to molecular physics \cite{Spradlin:1999bn,Cadoni:2000iz,Camblong:2003mz,Strominger:2003tm,Gaiotto:2004ij,Camblong:2004ye,Andrzejewski:2011ya,Jackiw:2012ur,Okazaki:2015pfa,Andrzejewski:2015jya,Khodaee:2017tbk,Okazaki:2017lpn,Tada:2017wul,BenAchour:2019ufa,Camblong:2022oet,Dorey:2022cfn,Dorey:2022ics,Dorey:2023jfw}. A major motivation to study and bootstrap CQM stems from the AdS/CFT correspondence \cite{Maldacena:1997re}. AdS$_2$/CFT$_1$ is often referred to as the runt of the correspondence as CFT$_1$ does not possess a local stress tensor that generates conformal transformations which is in sharp contrast to its higher dimensional counterparts. In the context of the DFF model, this correspondence and its subtleties were first discussed and explored in \cite{Chamon:2011xk}. Another extremely important case is that of the SYK models and their holographic bulk duals \cite{Maldacena:2016hyu,Gross:2017aos,Rosenhaus:2018dtp}. To investigate such dualities, having a firm foothold on the conformal theory side of things is essential. Most of the analysis of CQM has taken place in the real space aka, the time domain. However, in the past decade or so it has been discovered that working in momentum space leads to new insights and connections such as those between conformal field theory correlators and scattering amplitudes in one higher dimension, double copy relations and much more \cite{Coriano:2013jba,Bzowski:2013sza,Bzowski:2015pba,Bzowski:2017poo,Farrow:2018yni,Bzowski:2018fql,Bautista:2019qxj,Lipstein:2019mpu,Jain:2020rmw,Jain:2020puw,Jain:2021wyn,Jain:2021qcl,Jain:2021vrv,Jain:2021gwa,Jain:2021whr,Isono:2019ihz,Gillioz:2019lgs,Baumann:2019oyu}. Despite all this success, there has been a difficulty in setting up the conformal bootstrap intrinsically in momentum space. Although momentum space conformal blocks \cite{Gillioz:2018mto,Gillioz:2019iye,Gillioz:2020wgw} and even a simplex representation for generic scalar $n$-point functions \cite{Bzowski:2019kwd,Bzowski:2020kfw,Caloro:2022zuy} have been obtained, they are quite complicated to use and obtain interesting results. This is where the one-dimensional case can serve as a toy model, which one can then attempt to emulate in higher dimensions. Due to the decreased technical difficulties, one can probe deeper into the structure of momentum space CFT and obtain illuminating results. As a bonus, the conformal Ward identities in conformal quantum mechanics coincide with those in the  (anti-)holomorphic sector of two dimensional CFT \cite{Belavin:1984vu}, thereby making our results applicable there too. Thus, our aim in this paper is to set the stage for the momentum space CQM bootstrap.\\\\
\textbf{Outline:}\\
In section \ref{sec:stage}, we discuss the $\mathfrak{sl}(2,\mathbb{R})$ symmetry of conformal quantum mechanics and its implications on correlation functions of primary operators. In section \ref{sec:CQMcorrelators}, we follow up by solving the constraints due to the conformal invariance in momentum space. We obtain the general form of three point and four point functions, which we then generalize to arbitrary $n$-point functions. This analysis unveils the momentum space analog of the conformal cross ratios. We then follow this by the computation of momentum space conformal partial waves. We also provide an interesting momentum mellin space represention for the correlators. Further, we show that the presence of multiple solutions to the conformal Ward identities can be attributed to the various possible time orderings of the correlator. Moreover, we provide explicit checks of our formulae by free theory and DFF model computations. In sections \ref{sec:Neq1SCQMcorrelators} and \ref{sec:Neq2SCQMcorrelators}, we extend our analysis to $\mathcal{N}=1,2$ superconformal quantum mechanics and show that the supersymmetry constraints are identical to identities obeyed by hypergeometric functions. Finally, we discuss possible future directions in section \ref{sec:Discussion}. We provide appendices to supplement our main text. In appendices \ref{appendix:Neq1} and \ref{appendix:Neq2}, we present the symmetry algebra as well as the associated Ward identities for correlators in $\mathcal{N}=1,2$ SCQM. In appendix \ref{appendix:SeriesFormulae}, we provide series expansions for the Lauricella function of $n$ variables. We also provide some useful identities involving hypergeometric functions in appendix \ref{appendix:hypergeometric}.

\section{Setting the Stage}\label{sec:stage}
In this section, we lay the foundations for the remainder of the paper. We begin by discussing the symmetry algebra of conformal quantum mechanics. We then present the action of its generators on primary operators both in the time and energy representations. We then discuss the implications of the conformal ward identities on $n$ point correlation functions.

Conformal quantum mechanics can be viewed as a conformal field theory living in one dimension. In this paper, we work in the one dimensional manifold $M=\mathbb{R}$. Our metric reads,
\begin{align}
    ds^2=dt^2.
\end{align}
Under a general diffeomorphism $t\to f(t)$, we see that the metric is invariant upto an overall conformal factor:
\begin{equation}
    ds^2\to d\Tilde{s}^2=\left(\frac{df}{dt}\right)^2 dt^2=\Omega^2(t)ds^2.
\end{equation}
Therefore, the one dimensional conformal group is the infinite dimensional group $\mathbf{Diff}(\mathbb{R})$. The generators of this group which we denote by $l_n$, obey the infinite dimensional Witt algbera:
\begin{align}
    & l_n=t^{1-n}\frac{d}{dt}~~,~~[l_n,l_m]=(n-m)l_{n+m}.
\end{align}
However, $l_{-1}$, $l_0$ and $l_1$ are the only generators that are non singular on the entire real line. These generators form a closed $\mathfrak{sl}(2,\mathbb{R})$ sub-algebra and for the remainder of the paper, we restrict our study to it . This algebra is isomorphic to $\mathfrak{so}(2,1)$ which is in accordance with the fact that the conformal algebra in $d$ dimensions is $\mathfrak{so}(d+1,1)$ and hence $\mathfrak{so}(2,1)$ in $d=1$.

We henceforth denote $l_0,l_1$ and $l_{-1}$ as $H,D$ and $K$ respectively. $H$ is the Hamiltonian, $D$ is the dilatation operator and $K$ is the generator of special conformal transformations. The $\mathfrak{sl}(2,\mathbb{R})$ algebra that they obey is given by,
\begin{align}\label{algebra}
\begin{split}
    &[D,H]=-iH,\\
    &[D,K]=iK,\\
    &[K,H]=-2iD.
    \end{split}
\end{align}
The action of these generators on primary operators\footnote{Local operators placed at the origin transform in irreducible representations of $\mathfrak{sl}(2,\mathbb{R})$. An operator $\mathcal{O}_{\Delta}(t)$ is said to be primary if it satisfies, $[D,\mathcal{O}_{\Delta}(0)]=-i\Delta \mathcal{O}_{\Delta}(0)$ and $[K,\mathcal{O}_{\Delta}(0)]=0$ (or equivalently $[H,\mathcal{O}_{\Delta}(\infty)]=0$).} is given by,
\begin{equation}\label{HDKactiontspace}
    \begin{split}
          [H,\mathcal{O}_{\Delta}(t)]&=i\frac{d}{dt}\mathcal{O}_{\Delta}(t),\\
    [D,\mathcal{O}_{\Delta}(t)]&=i\left(t\frac{d}{dt}+\Delta\right)\mathcal{O}_{\Delta}(t),\\
[K,\mathcal{O}_{\Delta}(t)]&=i\left(t^2\frac{d}{dt}+2t\Delta\right)\mathcal{O}_{\Delta}(t).
    \end{split}
\end{equation}
It is straightforward to Fourier transform these expressions to obtain their momentum space (which in this case is just the energy or frequency space) counterparts:
\begin{equation}\label{HDKactionomegaspace}
\begin{split}
    [H,\mathcal{O}_{\Delta}(\omega)]&=\omega \mathcal{O}_{\Delta}(\omega),\\
    [D,\mathcal{O}_{\Delta}(\omega)]&=-i\left(\omega\frac{d}{d\omega}+(1-\Delta)\right)\mathcal{O}_{\Delta}(\omega),\\
    [K,\mathcal{O}_{\Delta}(\omega)]&=-\left(\omega\frac{d^2}{d\omega^2}+2(1-\Delta)\frac{d}{d\omega}\right)\mathcal{O}_{\Delta}(\omega).
    \end{split}
\end{equation}
We now consider the implications of this $\mathfrak{sl}(2,\mathbb{R})$ invariance on correlation functions of primary operators\footnote{As noted in \cite{Chamon:2011xk}, there are subtleties related to the $\mathfrak{sl}(2,\mathbb{R})$ conformal invariance of correlation functions in conformal quantum mechanics. However, at the end of the day, the authors show that an interplay of many non-trivial effects ensures this invariance. Therefore, we assume throughout this paper that we can define correlation functions appropriately such that they enjoy conformal invariance.}.

Consider an arbitrary $n$-point function of primary operators,
\begin{align}
    f_n(t_1,\cdots,t_n)=\langle O_{\Delta_1}(t_1)\cdots O_{\Delta_n}(t_n)\rangle.
\end{align}
The invariance of this correlator under the simultaneous action of $H,D$ or $K$ on all insertions, a.k.a the conformal ward identities are given by,
\begin{align}\label{tspaceWT1}
    &\langle [\mathcal{L},O_{\Delta_1}(t_1)]\cdots O_{\Delta_n}(t_n)\rangle+\cdots \langle O_{\Delta_1}(t_1)\cdots [\mathcal{L},O_{\Delta_n}(t_n)]\rangle=0~~,~~\mathcal{L}\in\{H,D,K\}.
\end{align}
Using the commutators provided in \eqref{HDKactiontspace} in \eqref{tspaceWT1} we obtain the following Ward identities, due to $H,D$ and $K$ respectively:
\begin{align}\label{tgenerators}
\begin{split}
    \sum_{i=1}^{n}\frac{\partial}{\partial t_i}f_n(t_1,\cdots,t_n)&=0,\\
\sum_{i=1}^{n}\left(t_i\frac{\partial}{\partial t_i}+\Delta_i\right)f_n(t_1,\cdots,t_n)&=0,\\
\sum_{i=1}^{n}\left(t_i^2\frac{\partial}{\partial t_i}+2 t_i \Delta_i\right)f_n(t_1,\cdots,t_n)&=0.
    \end{split}
\end{align}
The Fourier space counterparts of above equations are readily obtained and are given by,
\begin{align}
\sum_{i=1}^{n}\omega_if_n(\omega_1,\cdots,\omega_n)&=0,\label{omegaHgenerator}\\
    \sum_{i=1}^{n}\left(\omega_i\frac{\partial}{\partial \omega_i}+(1-\Delta_i)\right)f_n(\omega_1,\cdots,\omega_n)&=0,\label{omegaDgenerator}\\
    \sum_{i=1}^{n}\left(\omega_i\frac{\partial^2}{\partial \omega_i^2}+2(1-\Delta_i)\frac{\partial}{\partial\omega_i}\right)f_n(\omega_1,\cdots,\omega_n)&=0.\label{omegaKgenerator}
\end{align}
With the conformal ward identities, \eqref{tgenerators}, and \eqref{omegaHgenerator},\,\eqref{omegaDgenerator},\,\eqref{omegaKgenerator} in hand, we now proceed to solve them in the next section.
\section{Correlators in Conformal Quantum Mechanics}\label{sec:CQMcorrelators}
The task of obtaining closed form expressions in general dimensions for momentum space four and higher point functions has been quite difficult. So far, a simplex integral representation has been achieved for arbitrary scalar $n$-point functions \cite{Bzowski:2019kwd,Bzowski:2020kfw,Caloro:2022zuy}.
In this section, we shall show that we obtain a stronger result in one dimension, i.e., find closed form analytic expressions for not only four-point functions but arbitrary $n$-point ones.
We also see that expressing our results in a  Mellin-Barnes integral representation unveils an interesting structure for the Momentum-Mellin amplitude. Further, we shall compute conformal partial waves which we test against several examples. Moreover, we explain the existence of multiple solutions to the momentum space Conformal Ward identities as arising due to the in-equivalent Fourier transforms. We supplement and check our analysis by performing and comparing with examples in the free bosonic, free fermionic and DFF model.
\subsection{Solutions to the Conformal Ward Identities}\label{ConfWardSol}
Let us first present the general solution to two, three, and four point functions to the time domain Ward identities. By solving the differential equations \eqref{tgenerators} we obtain,
\begin{align}\label{234pointtspace}
    \langle O_{\Delta_1}(t_1)O_{\Delta_2}(t_2)\rangle&=\frac{c_{12}\delta_{\Delta_1,\Delta_2}}{|t_1-t_2|^{2\Delta}},\notag\\
    \langle O_{\Delta_1}(t_1)O_{\Delta_2}(t_2)O_{\Delta_3}(t_3)\rangle&=\frac{f_{123}}{|t_1-t_2|^{\Delta_1+\Delta_2-\Delta_3}|t_2-t_3|^{\Delta_2+\Delta_3-\Delta_1}|t_1-t_3|^{\Delta_1+\Delta_3-\Delta_2}},\notag\\
    \langle O_{\Delta_1}(t_1)O_{\Delta_2}(t_2)O_{\Delta_3}(t_3)O_{\Delta_4}(t_4)\rangle&=\prod_{1\le i\le j\le 4}(|t_i-t_j|)^{\frac{\Delta_1+\Delta_2+\Delta_3+\Delta_4}{3}-\Delta_i-\Delta_j}~G(\chi),\notag\\
    \langle O_{\Delta_1}(t_1)\cdots O_{\Delta_n}(t_n)\rangle&=\prod_{1\le i\le j\le n}(|t_i-t_j|)^{2\alpha_{ij}}G_n(\chi_1,\cdots \chi_{n-3}),
\end{align}
where, $\displaystyle \chi=\frac{|t_1-t_2||t_3-t_4|}{|t_1-t_3||t_2-t_4|}$ is the four point cross ratio, $\chi_1,\cdots \chi_{n-3}$ are the cross ratios for $n$-point functions which take the form $\displaystyle 
 \frac{|t_i-t_j||t_k-t_l|}{|t_i-t_k||t_j-t_l|},~i,j,k,l\in\{1,\cdots n\}$ and the $\alpha_{ij}$ satisfy $\displaystyle \Delta_i=-\sum_{j=1}^{n}\alpha_{ij}~,i\in\{1,\cdots,n\}$. An artefact of the fact that we are in one dimension is that we have only a single (real) cross ratio at the level of four points in contrast to the case in higher dimensions. For $n$ point functions, we have only $n-3$ cross ratios in contrast to the $\frac{n(n-3)}{2}$ cross ratios in sufficiently high dimensions.
 
Let us now systematically solve the momentum space conformal Ward identities \eqref{omegaHgenerator},\,\eqref{omegaDgenerator},\,\eqref{omegaKgenerator}  to obtain the analogue of \eqref{234pointtspace}.
\subsubsection{\normalsize{Two Point Functions}}
The translation ward identity \eqref{omegaHgenerator} entails the following form for the two point correlator:
\begin{align}
    &\langle O_{\Delta_1}(\omega_1)O_{\Delta_2}(\omega_2)\rangle=\delta(\omega_1+\omega_2)\Tilde{G}(\omega_1).
\end{align}
The dilatation ward identity \eqref{omegaDgenerator} yields,
\begin{align}
    \left(\omega_1\frac{\partial}{\partial\omega_1}+\omega_2\frac{\partial}{\partial\omega_2}+2-\Delta_1-\Delta_2\right)\big(\delta(\omega_1+\omega_2)\Tilde{G}(\omega_1)\big)=0.
\end{align}
We integrate both sides of this equation with respect to $\omega_2$ to obtain an equation to solve for $\Tilde{G}(\omega_1)$:
\begin{equation}\label{2pointGtfromD}
    \omega_1\frac{d\Tilde{G}(\omega_1)}{d\omega_1}=(\Delta_1+\Delta_2-1)\Tilde{G}(\omega_1)\implies \Tilde{G}(\omega_1)=\Tilde{C}_{12}\omega_1^{\Delta_1+\Delta_2-1}.
\end{equation}
Finally, the special conformal ward identity \eqref{omegaKgenerator} reads,
\begin{align}
\left(\omega_1\frac{\partial^2}{\partial\omega_1^2}+\omega_2\frac{\partial^2}{\partial\omega_2^2}+2(1-\Delta_1)\frac{\partial}{\partial\omega_1}+2(1-\Delta_2)\frac{\partial}{\partial\omega_2}\right)\big(\delta(\omega_1+\omega_2)\Tilde{G}(\omega_1)\big)&=0\notag\\
  \implies  \left(\omega_1\frac{d^2}{d\omega_1^2}+2(1-\Delta_1)\frac{d}{d\omega_1}\right) \Tilde{G}(\omega_1)=0\implies\left(\omega_1\frac{d^2}{d\omega_1^2}+2(1-\Delta_1)\frac{d}{d\omega_1}\right)\Tilde{C}_{12}\omega_1^{\Delta_1+\Delta_2-1}&=0\notag\\
\implies (\Delta_2-\Delta_1)(\Delta_1+\Delta_2-1)\Tilde{C}_{12}&=0.
\end{align}
where to go from the first to the second line, we integrated with respect to $\omega_2$ and from the second line we used \eqref{2pointGtfromD} which yields the result in the third line.
Therefore, we see that for operators with generic scaling dimensions, we require $\Delta_1=\Delta_2$ thus yielding,
\begin{align}\label{2pointSol}
    \langle O_{\Delta_1}(\omega_1)O_{\Delta_2}(\omega_2)\rangle=\Tilde{C}_{12}\delta_{\Delta_1,\Delta_2}\omega_1^{2\Delta_1-1}\delta(\omega_1+\omega_2),
\end{align}
which is indeed the Fourier transform of the time domain two point function provided in \eqref{234pointtspace} as can be easily verified. 
Let us now move on to the three point level.
\subsubsection{\normalsize{Three Point Functions}}
Translation invariance \eqref{omegaHgenerator} constrains the three point function to take the following form:
\begin{equation}
    \langle \mathcal{O}_{\Delta_1}(\omega_1)\mathcal{O}_{\Delta_2}(\omega_2)\mathcal{O}_{\Delta_3}(\omega_3)\rangle=\delta(\omega_1+\omega_2+\omega_3)\Tilde{G}(\omega_1,\omega_2).
\end{equation}
The dilatation ward identity \eqref{omegaDgenerator} yields,
 \begin{align}\label{3pointDWardId}
     \left(\omega_1\frac{\partial}{\partial\omega_1}+\omega_2\frac{\partial}{\partial\omega_2}+\omega_3\frac{\partial}{\partial\omega_3}+(3-\Delta_t)\right)\big(\delta(\omega_1+\omega_2+\omega_3)\Tilde{G}(\omega_1,\omega_2)\big)&=0\notag\\
   \implies \left(2-\Delta_t+\omega_1\frac{\partial}{\partial\omega_1}+\omega_2\frac{\partial}{\partial\omega_2}\right)\Tilde{G}(\omega_1,\omega_2)&=0.
 \end{align}
 where we defined $\Delta_t=\Delta_1+\Delta_2+\Delta_3$.\\
The special conformal ward identity \eqref{omegaKgenerator} demands,
\small
\begin{align}\label{3pointKWardId}
     \left(\omega_1\frac{\partial^2}{\partial\omega_1^2}+\omega_2\frac{\partial^2}{\partial\omega_2^2}+\omega_3\frac{\partial^2}{\partial\omega_3^2}+2(1-\Delta_1)\frac{\partial}{\partial\omega_1}+2(1-\Delta_2)\frac{\partial}{\partial\omega_2})+2(1-\Delta_3)\frac{\partial}{\partial\omega_3}\right)&\big(\delta(\omega_1+\omega_2+\omega_3)\Tilde{G}(\omega_1,\omega_2)\big)=0\notag\\
     \implies \left(\omega_1\frac{\partial^2}{\partial\omega_1^2}+\omega_2\frac{\partial^2}{\partial\omega_2^2}+2(1-\Delta_1)\frac{\partial}{\partial\omega_1}+2(1-\Delta_2)\frac{\partial}{\partial\omega_2}\right)\Tilde{G}(\omega_1,\omega_2)&=0.
\end{align}
\normalsize
Therefore, our task is to simultaneously solve the PDEs in \eqref{3pointDWardId} and \eqref{3pointKWardId} for the function $\Tilde{G}(\omega_1,\omega_2)$.
The solution to the dilatation ward identity \eqref{3pointDWardId} is easily obtained:
\begin{equation}\label{3pointDWardIdsol}
    \Tilde{G}(\omega_1,\omega_2)=\omega_1^{\Delta_t-2}\Tilde{g}(x),~~x=\frac{\omega_2}{\omega_1}.
\end{equation}
Plugging \eqref{3pointDWardIdsol} into \eqref{3pointKWardId} results in an ordinary differential equation for $\Tilde{g}(x)$:
\begin{align}\label{3pointDiffeqforg}
  &(x^2+x)\frac{d^2 \Tilde{g}(x)}{d x^2}-2\big(\Delta_2-1+x(\Delta_t-\Delta_1-2)\big)\frac{d \Tilde{g}(x)}{dx}+(\Delta_t-2)(\Delta_t-2\Delta_1-1)\Tilde{g}(x)=0,
\end{align}
which we recognize as the hypergeometric differential equation. We then solve \eqref{3pointDiffeqforg} to obtain the most general solution:
\begin{align}\label{3pointSolforg}
      \tilde{g}(x)&=c_1 ~_2F_1(2-\Delta_t,1+2\Delta _1-\Delta_t,2-2 \Delta _2;-x)
    \notag\\&\quad\qquad+c_2 \,x^{2\Delta_2-1}  ~_2F_1(1+2\Delta_2-\Delta_t,\Delta_t-2\Delta_3,2 \Delta _2;-x),~~c_1,c_2\in\mathbb{R}.
\end{align}
Therefore, the three point function in $\omega$ space is given by,
\begin{align}\label{3pointSol}
    \langle O_{\Delta_1}(\omega_1)O_{\Delta_2}(\omega_2)O_{\Delta_3}(\omega_3)&\rangle=\omega_1^{\Delta_1+\Delta_2+\Delta_3-2}\bigg[c_1 ~_2F_1(2-\Delta_t,1+2\Delta _1-\Delta_t,2-2 \Delta _2;-x)
    \notag\\&+c_2\, x^{2\Delta_2-1}  ~_2F_1(1+2\Delta_2-\Delta_t,\Delta_t-2\Delta_3,2 \Delta _2;-x) \bigg]\delta(\omega_1+\omega_2+\omega_3).
\end{align}
In contrast to the unique three point correlator in the time domain provided in  \eqref{234pointtspace}, we have here, two linearly independent solutions. The interpretation of the same will be made clear in subsection \ref{sec:TimeOrderingStuff} where we will show that the two solutions correspond to the Fourier transform of the various possible ``time orderings" of the time domain three point function. Let us also note that \eqref{3pointSol} is consistent with the results obtained in \cite{Gillioz:2019iye} for the momentum space three point functions in the holomorphic correlators in two dimensional CFT.
\subsubsection{\normalsize{Four Point Functions}}
We now repeat the same drill that we carried out in the analysis of two and three point functions for the four point case.\\ Translation invariance \eqref{omegaHgenerator} instills the following form for the correlator:
\begin{align}
    \langle O_{\Delta_1}(\omega_1)O_{\Delta_2}(\omega_2)O_{\Delta_3}(\omega_3)O_{\Delta_4}(\omega_4)\rangle=\delta(\omega_1+\omega_2+\omega_3+\omega_4)\Tilde{G}(\omega_1,\omega_2,\omega_3).
\end{align}
The dilatation ward identity \eqref{omegaDgenerator} yields,
\begin{align}
\bigg(\omega_1\frac{\partial}{\partial\omega_1}+\omega_2\frac{\partial}{\partial\omega_2}+\omega_3\frac{\partial}{\partial\omega_3}+(3-\Delta_t)\bigg)\Tilde{G}(\omega_1,\omega_2,\omega_3)&=0\notag\\
\implies \Tilde{G}(\omega_1,\omega_2,\omega_3)&=\omega_1^{\Delta_t-3}\Tilde{g}(x,y),
\end{align}
where we have defined $\Delta_t=\Delta_1+\Delta_2+\Delta_3+\Delta_4$ and the ratios $\displaystyle x=\frac{\omega_2}{\omega_1}$ and $\displaystyle y=\frac{\omega_3}{\omega_1}$.

The ward identity due to special conformal transformations \eqref{omegaKgenerator} gives us the following partial differential that the function $\Tilde{g}(x,y)$ has to satisfy:
\begin{align}\label{4pointKWardId}
    &\bigg(x(x+1)\frac{\partial^2}{\partial x^2}+y(y+1)\frac{\partial^2}{\partial y^2}+2 xy\frac{\partial^2}{\partial x\partial y}-2\big(\Delta_2-1+x(\Delta_t-\Delta_1-4)\big)\frac{\partial}{\partial x}\notag\\
    &-2\big(\Delta_3-1+y(\Delta_t-\Delta_1-4)\big)\frac{\partial}{\partial y}+(\Delta_t-3)(\Delta_t-2\Delta_1-2)\bigg)\Tilde{g}(x,y)=0.
\end{align}
Very interestingly, \eqref{4pointKWardId} is an equation obeyed by none other than Appell's generalized hypergeometric function $F_2$ \cite{AppellOG}! The system of differential equations that $F_2(a,b_1,b_2,c_1,c_2;x,y)$ obeys is the following:
\begin{align}
    &x(1-x)\frac{\partial^2F_2}{\partial x^2}-x y\frac{\partial^2 F_2}{\partial x\partial y}+(c_1-(a+b_1+1)x)\frac{\partial F_2}{\partial x}-b_1 y\frac{\partial F_2}{\partial y}-ab_1 F_2=0,\notag\\
    &y(1-y)\frac{\partial^2F_2}{\partial y^2}-x y\frac{\partial^2 F_2}{\partial x\partial y}+(c_2-(a+b_2+1)y)\frac{\partial F_2}{\partial y}-b_2 x\frac{\partial F_2}{\partial x}-ab_2 F_2=0.
\end{align}
If we add these two equations we obtain,
\small
\begin{align}\label{AppellF2eqnsum}
    &x(1-x)\frac{\partial^2 F_2}{\partial x^2}+(c_1-(a+b_1+b_2+1)x)\frac{\partial F_2}{\partial x }+y(1-y)\frac{\partial^2 F_2}{\partial y^2}+(c_2-(a+b_1+b_2+1)y)\frac{\partial F_2}{\partial y }-2xy\frac{\partial^2 F_2}{\partial x \partial y}-a(b_1+b_2) F_2=0.
\end{align}
\normalsize
Let us now perform the following re-labeling and mapping:
\begin{align}\label{relabelling}
 &x\to -x,\quad y\to -y,\qquad c_1=2(1-\Delta_2),\quad c_2=2(1-\Delta_3),\notag\\
    &\big\{a=(3-\Delta_t),~~\sum_{i=1}^{2}b_i=2+2\Delta_1-\Delta_t\big\}~\text{or}~\big\{a=2+2\Delta_1-\Delta_t,~~\sum_{i=1}^{2}b_i=3-\Delta_t\big\}.
\end{align}
Thus, \eqref{AppellF2eqnsum} becomes identical to our equation for the four point function \eqref{4pointKWardId}! We see that \eqref{relabelling} fixes $a,c_1$, $c_2',b_1+b_2$ in terms of the scaling dimensions of the external operators. The key point here is that the combination $b_1+b_2$ is fixed in terms of the external operator scaling dimensions but not $b_1$ and $b_2$ individually. If we fix $b_2$ in terms of $b_1$ using \eqref{relabelling}, then $b_1$ is left completely undetermined. Thus, our solution to the four-point function is\footnote{There are four linearly independent solutions to Appell's differential equation for $F_2$. Since $b_1$ is not determined via conformal invariance, we allow generic four-point functions to be a linear combination of solutions with different values of $b_1$. The constants $k_1,\cdots,k_4$ can depend on $b_1$ but for the sake of brevity, we suppress this dependence.},
\small
\begin{align}\label{4pointSol}
    \langle O_{\Delta_1}(\omega_1)O_{\Delta_2}(\omega_2)O_{\Delta_3}(\omega_3)O_{\Delta_4}(\omega_4)\rangle&=\sum_{b_1}\omega_1^{\Delta_t-3}\delta(\omega_1+\omega_2+\omega_3+\omega_4)\bigg[k_1F_2(a,b_1,b_2,c_1,c_2;-x,-y)\notag\\
    &+k_2(-x)^{1-c_1}F_2(a-c_1+1,b_1-c_1+1,b_2,2-c_1,c_2;-x,-y)\notag\\
    &+k_3(-y)^{1-c_2}F_2(a-c_2+1,b_1,b_2-c_2+1,c_1,2-c_2;-x,-y)\notag\\
    &+k_4(-x)^{1-c_1}(-y)^{1-c_2}F_2(a-c_1-c_2+2,b_1-c_1+1,b_2-c_2+1,2-c_1,2-c_2;-x,-y)\bigg],
\end{align}
with the parameters given in \eqref{relabelling}. A series expansion for $F_2$ is provided in appendix \ref{appendix:SeriesFormulae}. Contrasting this expression with the time domain four-point function in \eqref{234pointtspace} where we had a single undetermined function of $\chi$ (the cross-ratio), here we have an Appell $F_2$ function\footnote{Strictly speaking, we have four solutions corresponding to the four solutions of Appell's $F_2$ differential equation. The existence of multiple solutions rather than a unique one is also what we found at the three point level \eqref{3pointSol}. As mentioned earlier, we will elaborate on this in subsection \ref{sec:TimeOrderingStuff}.} with one undetermined parameter $b_1$ which shows that $b_1$ is the momentum space counterpart to the cross-ratio $\chi$. However, some comments are in order. Although \eqref{4pointSol} solves the conformal ward identities, it may not be the unique solution. The reason for the same is that the Appell $F_2$ functions are the unique solutions to the two equations in \eqref{AppellF2eqnsum}. Our equation \eqref{AppellF2eqnsum} is the sum of these two equations, and obviously, while the Appell $F_2$ functions are solutions, they may not be the unique ones. However, every explicit computation we performed to compute a four-point function is in accordance to \eqref{4pointSol}. If there exists any other solution to the conformal ward identities, there may be some additional consistency criteria\footnote{For example, OPE consistency was used in \cite{Jain:2021whr} to rule out certain solutions to the conformal ward identities.} to remove them. We defer such an analysis to future work. Let us move on to the even more complicated case of five-point functions.
\subsubsection{Five Point Functions}
By solving the translation and dilatation Ward identities (\eqref{omegaHgenerator} and \eqref{omegaDgenerator} respectively) we see that the correlator takes the following form:
\begin{align}
    \langle O_{\Delta_1}(\omega_1)O_{\Delta_2}(\omega_2)O_{\Delta_3}(\omega_3)O_{\Delta_4}(\omega_4)O_{\Delta_5}(\omega_5)\rangle=\delta(\omega_1+\omega_2+\omega_3+\omega_4+\omega_5)\omega_1^{\Delta_t-4}\Tilde{g}(x,y,z),
\end{align}
where we have defined,
\begin{align}
\Delta_t=\sum_{i=1}^{5}\Delta_i~,~x=\frac{\omega_2}{\omega_1},y=\frac{\omega_3}{\omega_1},z=\frac{\omega_4}{\omega_1}.
\end{align}
The special conformal Ward identity \eqref{omegaKgenerator} implies that the function $\Tilde{g}(x,y,z)$ satisfies the following differential equation:
\begin{align}\label{5pointSCTid}
    &\bigg(x(x+1)\frac{\partial^2}{\partial x^2}+y(y+1)\frac{\partial^2}{\partial y^2}+z(z+1)\frac{\partial^2}{\partial z^2}+2 xy\frac{\partial^2}{\partial x\partial y}+2yz\frac{\partial^2}{\partial y \partial z}+2 z x\frac{\partial^2}{\partial z \partial x}-2\big(\Delta_2-1+x(\Delta_t-\Delta_1-4)\big)\frac{\partial}{\partial x}\notag\\
    &-2\big(\Delta_3-1+y(\Delta_t-\Delta_1-4)\big)\frac{\partial}{\partial y}-2\big(\Delta_4-1+z(\Delta_t-\Delta_1-4)\big)\frac{\partial}{\partial z}+(\Delta_t-4)(\Delta_t-2\Delta_1-3)\bigg)\Tilde{g}(x,y,z)=0.
\end{align}
Just like we mapped the special conformal Ward identity for the four point function \eqref{4pointKWardId} to an equation obeyed by Appell's $F_2$ function, we were able to map \eqref{5pointSCTid} to an equation satisfied by the three variable Lauricella function $E^{(3)}_A$ \cite{LauricellaOG}! This three variable Lauricella function\footnote{ We provide a series expansion for $E_A^{(3)}$ in appendix \ref{appendix:SeriesFormulae}.} is the solution to the following system of three partial differential equations:
\begin{align}
    &\mathcal{L}^{(3)}_{i}E^{(3)}_{A}(a,b_1,b_2,b_3,c_1,c_2,c_3,x_1,x_2,x_3)=0,~i\in\{1,2,3\}~~\text{where},\notag\\
    &\mathcal{L}^{(3)}_{i}=x_i(1-x_i)\frac{\partial^2}{\partial x_i^2}-x_i\sum_{j\ne i}^{3}x_j\frac{\partial^2}{\partial x_i\partial x_j}+(c_i-(a+b_i+1)x_i)\frac{\partial}{\partial x_i}-b_i\sum_{j\ne i}^{3}x_j\frac{\partial}{\partial x_j}.
\end{align}
This obviously implies,
\begin{align}\label{3varLauriEQ1}
   &\sum_{i=1}^{3}\mathcal{L}_{i}^{(3)}E^{(3)}_{A}(a,b_1,b_2,b_3,c_1,c_2,c_3,x_1,x_2,x_3)=0.
\end{align}
If we choose,
\begin{align}\label{5pointmap}
    &x_1=-x,~~x_2=-y,~~x_3=-z,~~c_1=2(1-\Delta_2),~~c_2=2(1-\Delta_3),~~c_3=2(1-\Delta_4),\notag\\
    &\big\{a=(4-\Delta_t),~\sum_{i=1}^{3}b_i=3+2\Delta_1-\Delta_t\big\}~\text{or}~\big\{a=3+2\Delta_1-\Delta_t,~\sum_{i=1}^{3}b_i=4-\Delta_t\big\}.
\end{align}
Thus, \eqref{3varLauriEQ1} becomes identical to the five point conformal Ward identity \eqref{5pointSCTid}! Note that the map \eqref{5pointmap} specifies $c_1,c_2,c_3$ and $a$ in terms of the scaling dimensions of the external operators. As for the $b_i$, it fixes just one of them in terms of the other too, say, $b_3$ in terms of $b_1$, $b_2$. The number of five point cross ratios is two and the number of undetermined parameters we have is also two. Thus, we have found a momentum space analogue for these two cross ratios. Putting everything together, our solution for five point functions reads,
 \begin{align}
     \langle O_{\Delta_1}(\omega_1)O_{\Delta_2}(\omega_2)O_{\Delta_3}(\omega_3)O_{\Delta_4}(\omega_4)O_{\Delta_5}(\omega_5)\rangle=\delta(\omega_1&+\omega_2 +\omega_3+\omega_4+\omega_5+\omega_6)\omega_1^{\Delta_t-4}\notag\\
     &\big(k_1E_A^{(3)}(a,b_1,b_2,b_3,c_1,c_2,c_3,-x,-y,-z)+\cdots,
 \end{align}
 with the parameters provided in \eqref{5pointmap}. The dots indicate the seven other linearly independent solutions to the Lauricella $E_A^{(3)}$ PDE. We have also suppressed a possible sum over different values of $b_1$ and $b_2$ as they are not fixed by conformal invariance and a single correlator can be a linear combination of structures with different $b_1$ and $b_2$. With four and five point functions in hand, let us now see if we can make a similar statement for six point functions.
\subsubsection{Six Point Functions}
By solving the translation and dilatation Ward identities (\eqref{omegaHgenerator} and \eqref{omegaDgenerator} respectively) we see that the correlator takes the following form:
\begin{align}
    \langle O_{\Delta_1}(\omega_1)O_{\Delta_2}(\omega_2)O_{\Delta_3}(\omega_3)O_{\Delta_4}(\omega_4)O_{\Delta_5}(\omega_5)O_{\Delta_6}(\omega_6)\rangle=\delta(\omega_1+\omega_2+\omega_3+\omega_4+\omega_5+\omega_6)\omega_1^{\Delta_t-5}\Tilde{g}(x,y,z,u),
\end{align}
where we have defined,
\begin{align}
\Delta_t=\sum_{i=1}^{6}\Delta_i~,~x=\frac{\omega_2}{\omega_1},y=\frac{\omega_3}{\omega_1},z=\frac{\omega_4}{\omega_1},u=\frac{\omega_5}{\omega_1}.
\end{align}
The special conformal Ward identity \eqref{omegaKgenerator} implies that the function $\Tilde{g}(x,y,z,u)$ satisfies the following differential equation:
\begin{align}\label{6pointSCTid}
    &\bigg(x(x+1)\frac{\partial^2}{\partial x^2}+y(y+1)\frac{\partial^2}{\partial y^2}+z(z+1)\frac{\partial^2}{\partial z^2}+u(u+1)\frac{\partial^2}{\partial u^2}+2 xy\frac{\partial^2}{\partial x\partial y}+2yz\frac{\partial^2}{\partial y \partial z}+2 z x\frac{\partial^2}{\partial z \partial x}\notag\\&+2xu\frac{\partial^2}{\partial x\partial u}+2 y u\frac{\partial^2}{\partial y \partial u}+2 z u\frac{\partial^2}{\partial z\partial u}-2\big(\Delta_2-1+x(\Delta_t-\Delta_1-5)\big)\frac{\partial}{\partial x}-2\big(\Delta_3-1+y(\Delta_t-\Delta_1-5)\big)\frac{\partial}{\partial y}\notag\\
    &-2\big(\Delta_4-1+z(\Delta_t-\Delta_1-5)\big)\frac{\partial}{\partial z}-2\big(\Delta_5-1+u(\Delta_t-\Delta_1-5)\big)\frac{\partial}{\partial u}+(\Delta_t-5)(\Delta_t-2\Delta_1-4)\bigg)\Tilde{g}(x,y,z,u)=0.
\end{align}
A solution to \eqref{6pointSCTid} turns out to be none other than the four variable Lauricella function $E_{A}^{(4)}$ \cite{matsumoto}!
This four variable Lauricella function is the solution to the following system of four partial differential equations:
\begin{align}
    &\mathcal{L}^{(4)}_{i}E^{(4)}_{A}(a,b_1,b_2,b_3,b_4,c_1,c_2,c_3,c_4,x_1,x_2,x_3,x_4)=0,~i\in\{1,2,3\}~~\text{where},\notag\\
    &\mathcal{L}^{(4)}_{i}=x_i(1-x_i)\frac{\partial^2}{\partial x_i^2}-x_i\sum_{j\ne i}^{4}x_j\frac{\partial^2}{\partial x_i\partial x_j}+(c_i-(a+b_i+1)x_i)\frac{\partial}{\partial x_i}-b_i\sum_{j\ne i}^{4}x_j\frac{\partial}{\partial x_j}.
\end{align}
This obviously implies,
\begin{align}\label{4varLauriEQ1}
   &\sum_{i=1}^{3}\mathcal{L}_{i}^{(4)}E^{(4)}_{A}(a,b_1,b_2,b_3,b_4,c_1,c_2,c_3,c_4,x_1,x_2,x_3,x_4)=0.
\end{align}
If we choose,
\begin{align}\label{6pointmap}
    &x_1=-x,~~x_2=-y,~~x_3=-z,~~x_4=-u,~~c_1=2(1-\Delta_2),~~c_2=2(1-\Delta_3),~~c_3=2(1-\Delta_4),~~c_4=2(1-\Delta_5),\notag\\
    &\qquad\qquad\qquad\big\{a=(5-\Delta_t),\sum_{i=1}^{3}b_i=4+2\Delta_1-\Delta_t\big\}~\text{or}~\big\{a=4+2\Delta_1-\Delta_t,\sum_{i=1}^{3}b_i=5-\Delta_t\big\},
\end{align}
\eqref{4varLauriEQ1} becomes identical to \eqref{6pointSCTid}! Thus, our solution for six point functions is,
\begin{align}
    \langle O_{\Delta_1}(\omega_1)O_{\Delta_2}(\omega_2)O_{\Delta_3}(\omega_3)O_{\Delta_4}(\omega_4)O_{\Delta_5}(\omega_5)O_{\Delta_6}(\omega_6)\rangle&=\bigg(\delta(\omega_1+\omega_2+\omega_3+\omega_4+\omega_5+\omega_6)\omega_1^{\Delta_t-5}\notag\\
    &~~k_1 E_A^{(4)}(a,b_1,b_2,b_3,b_4,c_1,c_2,c_3,c_4;-x,-y,-z,-u)+\cdots\bigg),
\end{align}
where parameters are given in \eqref{6pointmap} and the $\cdots$ stand for the $15$ other linearly independent solutions to the Lauricella $E_A^{(4)}$ PDE \footnote{A series expansion for $E_A^{(4)}$ is provided in appendix \ref{appendix:SeriesFormulae}.}. Note that \eqref{6pointmap} fixes $a$ and the $c_i$ but leaves say, $b_1,b_2$ and $b_3$ undetermined. Thus the general six point function can be a sum over such solutions with different values of $b_1,b_2$ and $b_3$. Note that three is also the number of independent cross ratios for six point functions, indicating that these parameters are their momentum space analogue. Motivated by our success at the four, five, and six-point levels, let us attempt to generalize our results to arbitrary $n$-point functions.
\subsubsection{Generalization to $n$ point functions}
By solving the translation and dilatation Ward identities (\eqref{omegaHgenerator} and \eqref{omegaDgenerator} respectively) we see that any $n$ point correlator takes the following form:
\begin{align}
    \langle O_{\Delta_1}(\omega_1)\cdots O_{\Delta_n}(\omega_n)\rangle=\delta(\omega_1+\cdots+\omega_n)\omega_1^{\Delta_t-(n-1)}\Tilde{g}(y_1,\cdots,y_{n-2}),
\end{align}
where we have defined,
\begin{align}
\Delta_t=\sum_{i=1}^{n}\Delta_i~,~y_i=\frac{\omega_{i+1}}{\omega_1}.
\end{align}
The special conformal Ward identity \eqref{omegaKgenerator} implies that the function $\Tilde{g}(y_1,\cdots,y_{n-2})$ satisfies the following differential equation:
\begin{align}\label{npointSCTid}
    \bigg(\sum_{i=1}^{n-2}y_i(y_i+1)\frac{\partial^2}{\partial y_i^2}+& 2\sum_{1\le j<i\le n-2}y_i y_j\frac{\partial^2}{\partial y_i \partial y_j}-2\sum_{i=1}^{n-2}\big(\Delta_{i+1}-1+y_i(\Delta_t-\Delta_1-n+1)\big)\frac{\partial}{\partial y_i}\notag\\
    &+(\Delta_t-n+1)(\Delta_t-2\Delta_1-n+2)\bigg)\Tilde{g}(y_1,\cdots,y_{n-2})=0.
\end{align}
The generalized Lauricella function of $n-2$ variables, $E_A^{(n-2)}(a,b_1,\cdots, b_{n-2},c_1,\cdots,c_{n-2};x_1,\cdots, x_{n-2})$ obeys the following system of PDEs \cite{matsumoto}:
\begin{align}
    &\mathcal{L}^{(n-2)}_{i}E_A^{(n-2)}(a,b_1,\cdots,b_{n-2},c_1,\cdots,c_{n-2};x_1,\cdots, x_{n-2})=0,~i\in\{1,\cdots, n-2\},~\text{where},\notag\\
    &\mathcal{L}^{(n-2)}_{i}=x_i(1-x_i)\frac{\partial^2}{\partial x_i^2}-x_i\sum_{j\ne i}^{n-2}x_j\frac{\partial^2}{\partial x_i\partial x_j}+(c_i-(a+b_i+1)x_i)\frac{\partial}{\partial x_i}-b_i\sum_{j\ne i}^{n-2}x_j\frac{\partial}{\partial x_j}.
\end{align}
This obviously implies,
\begin{align}\label{nminus2varLauriEQ1}
   &\sum_{i=1}^{n-2}\mathcal{L}_{i}^{(n-2)}E_A^{(n-2)}(a,b_1,\cdots, b_{n-2},c_1,\cdots,c_{n-2};x_1,\cdots, x_{n-2})=0.
\end{align}
If we choose,
\begin{align}\label{npointmap}
&\qquad\qquad\qquad\qquad\qquad\qquad\qquad\qquad~~ x_i=-y_i,~~ c_i=2(1-\Delta_{i+1}),\notag\\
    &\big\{a=n-1-\Delta_t,~\sum_{i=1}^{n-2}b_i=n-2+2\Delta_1-\Delta_t\big\}~\text{or}~\big\{a=n-2+2\Delta_1-\Delta_t,~\sum_{i=1}^{n-2}b_i=n-1-\Delta_t\big\}.
\end{align}
Thus, \eqref{nminus2varLauriEQ1} becomes identical to the n point conformal Ward identity \eqref{npointSCTid}! The map \eqref{npointmap} fixes $a$ and the $n-2$ $c_i$ but leaves $n-3$ out of the $n-2$ $b_i$ undetermined. This is exactly the number of independent conformal cross ratios in the time domain and thus we have obtained their momentum space analogues for any n point correlation function. Our result for the $n$ point function reads,
\begin{align}\label{nPointSol}
     \langle O_{\Delta_1}(\omega_1)\cdots O_{\Delta_n}(\omega_n)\rangle=\delta(\omega_1+\cdots+\omega_n)\omega_1^{\Delta_t-(n-1)}\bigg(k_1 E_A^{(n-2)}(a,b_1,\cdots,b_{n-2},c_1,\cdots,c_{n-2};-x_1,\cdots,-x_{n-2})+\cdots\bigg),
\end{align}
where the parameters are given in \eqref{npointmap} and the $\cdots$ stand for the remaining $2^{n-2}$ linearly independent solutions to the Lauricella $E_A^{(n-2)}$ PDEs\footnote{It is interesting to note that certain Lauricella functions also pop out as solutions of conformal integrals \cite{Pal:2023kgu} as well as in solutions to the conformal Ward identity in special kinematics \cite{Coriano:2019nkw}.}(We provide a series expansion for $E_A^{(n-2)}$ in appendix \ref{appendix:SeriesFormulae}). As in the previous cases, we have suppressed a possible sum over the $b_i$ that are not determined by conformal invariance. The result \eqref{nPointSol} is quite pleasing and presents the first closed-form expression for a conformal $n$-point function in momentum space\footnote{Previously, a simplex representation for $n$ point functions in momentum space was obtained \cite{Bzowski:2019kwd}. Their result, in contrast to out algebraic formulae \eqref{nPointSol} is a simplex integral representation for the correlator.}.
\subsection{A Momentum-Mellin Space representation}
Motivated by the (generalized) hypergeometric structure of our correlators, we investigate their Mellin space representation. Consider the four point function that we obtained \eqref{4pointSol}. For simplicity we focus on the first term in the expression \eqref{4pointSol}\footnote{As will be explained in section \ref{sec:TimeOrderingStuff}, the different solutions represent the Fourier space counterparts of the in-equivalent time orderings.}. For the convenience of the reader we provide it here\footnote{We suppress the momentum conserving delta function $\delta(\omega_1+\omega_2+\omega_3+\omega_4)$ for brevity.}:
\begin{align}\label{G4}
    \Tilde{G}_4(\omega_1,\omega_2,\omega_3,\omega_4)=\sum_{b_1}k_1(b_1)\omega_1^{\Delta_t-3}F_2(a,b_1,b_2,c_1,c_2;-x,-y).
\end{align}
where,
\begin{align}\label{consts4point}
     c_1=2(1-\Delta_2),~~ c_2=2(1-\Delta_3),~~\big\{a=2+2\Delta_1-\Delta_t,~~\sum_{i=1}^{2}b_i=3-\Delta_t\big\},~~\Delta_t=\Delta_1+\Delta_2+\Delta_3+\Delta_4,
\end{align}
and $k_1(b_1)$ are constants that weigh the contribution of a particular $b_1$ to the correlator.
\\
Using a Mellin-Barnes type integral representation of $F_2(a,b_1,b_2,c_1,c_2;-x,-y)$ (for instance, see section 5.8.3 in \cite{bateman1953higher}), we obtain the following beautiful form for the correlator \eqref{G4}:
\begin{align}\label{MellinMomSpaceAmplitudeFormula}
\Tilde{G}_4(\omega_1,\omega_2,\omega_3,\omega_4)=\omega_1^{\Delta_t-3}\frac{\Gamma(2-2\Delta_2)\Gamma(2-2\Delta_3)}{\Gamma(3-\Delta_t)}\int_{-i \infty}^{i \infty}\frac{ds}{2\pi i}\int_{-i \infty}^{i\infty}\frac{dt}{2\pi i}\frac{\Gamma(s+t+2+2\Delta_1-\Delta_t)\Gamma(-s)\Gamma(-t)}{\Gamma(s+2-2\Delta_2)\Gamma(t+2-2\Delta_3)}x^s y^t\mathcal{M}(s,t),
\end{align}
where we have defined the momentum-Mellin space amplitude,
\begin{align}\label{MellinMomSpace}
    \mathcal{M}(s,t)=\sum_{b_1}k_1(b_1)\mathcal{M}_{b_1}(s,t)=\sum_{b_1}k_1(b_1)\frac{\Gamma(b_1+s)\Gamma(t+3-\Delta_t-b_1)}{\Gamma(b_1)\Gamma(3-\Delta_t-b_1)}.
\end{align}
where $\mathcal{M}_{b_1}(s,t)$ are momentum-Mellin space partial amplitudes which represent the contribution of a particular $b_1$ to the full amplitude. The pre-factors in \eqref{MellinMomSpaceAmplitudeFormula} are fixed by the external dimensions of the operators (kinematics) whereas $\mathcal{M}(s,t)$ represents the dynamical contributions to the correlator.
Note the contrast to the usual real space Mellin amplitudes \cite{Penedones:2010ue,Gopakumar:2016cpb}: The usual Mellin amplitude is an \textit{unknown} (theory dependent) function of $s$ and $t$. Here in \eqref{MellinMomSpace}, the theory dependence just lies in the constants $k_1(b_1)$ which weighs the contribution of the different $b_1$ to the correlator. The functional form of the amplitude, however, is completely determined by the above ratio of gamma functions.

For example, the momentum space correlator $\langle O_1O_1O_1O_1\rangle$ that we compute in \eqref{OOOObeta} in the free bosonic theory receives contributions only from $b_1=4,5$ and $6$ with $k_1(b_1)=14,28$ and $40$ respectively. Using these facts in \eqref{MellinMomSpace}, we obtain the following compact momentum-mellin amplitude:
\begin{align}\label{OOOOexamplemellinmomspace}
    \mathcal{M}(s,t)&=. \frac{7}{6}\Gamma(s+4)\Gamma(t+3)+\frac{7}{6}\Gamma(s+5)\Gamma(t+2)+\frac{1}{3}\Gamma(s+6)\Gamma(t+1)\notag\\
    &=\frac{\big(82+2s^2+7t(7+t)+s(25+7t)\big)}{6}\Gamma(4+s)\Gamma(1+t).
\end{align}
Based on the simplicity of this representation, it would be interesting to directly bootstrap four and higher point correlators in the momentum-mellin space. We leave this analysis to a future work. 
\subsection{Momentum space Conformal Partial Waves}
In this subsection, we compute momentum space conformal partial waves. Conformal partial waves are eigenvectors of the quadratic conformal Casimir $C_2$ \cite{Dolan:2003hv}. The expression for $C_2$ is as follows:
\begin{align}
    C_2=\frac{1}{2}(HK+KH)-D^2.
\end{align}
It is easy to see that $C_2$ commutes with $H,K$ and $D$ using the $\mathfrak{sl}(2,\mathbb{R})$ conformal algebra \eqref{algebra}. The conformal partial wave $W_{\Delta}(t_1,t_2,t_3,t_4)$ satisfies the following differential equation:
\begin{align}
    C_{12}W_{\Delta}(t_1,t_2,t_3,t_4)=\Delta(\Delta-1)W_{\Delta}(t_1,t_2,t_3,t_4).
\end{align}
The conformal partial wave (in the s channel) is given by the following integral \cite{Simmons-Duffin:2012juh}:
\begin{align}\label{conformalpartialwavedef}
    W^{(s)}_{\Delta}(t_1,t_2,t_3,t_4)=\int dt \langle O_{\Delta_1}(t_1)O_{\Delta_2}(t_2)O_{\Delta}(t)\rangle\langle\Tilde{O}_{\Delta}(t)O_{\Delta_3}(t_3)O_{\Delta_4}(t_4)\rangle,
\end{align}
where we have introduced the shadow operator, $\Tilde{O}$, which is defined as,
\begin{align}\label{shadowop}
    \Tilde{O}(t)=\int_{-\infty}^{\infty}~\frac{dx}{|x-t|^{2-2\Delta}}O_{\Delta}(t).
\end{align}
Using \eqref{shadowop} in \eqref{conformalpartialwavedef} we obtain,
\begin{align}
     W^{(s)}_{\Delta}(t_1,t_2,t_3,t_4)=\int_{-\infty}^{\infty}\int_{-\infty}^{\infty} \frac{dt dx}{|x-t|^{2-2\Delta}} \langle O_{\Delta_1}(t_1)O_{\Delta_2}(t_2)O_{\Delta}(t)\rangle\langle O_{\Delta}(t)O_{\Delta_3}(t_3)O_{\Delta_4}(t_4)\rangle.
\end{align}
Fourier transforming with respect to $t_1,t_2,t_3$ and $t_4$ we obtain,
\begin{align}
     W^{(s)}_{\Delta}(\omega_1,\omega_2,\omega_3,\omega_4)=\int_{-\infty}^{\infty}\int_{-\infty}^{\infty} \frac{dt dx}{|x-t|^{2-2\Delta}} \langle O_{\Delta_1}(\omega_1)O_{\Delta_2}(\omega_2)O_{\Delta}(t)\rangle\langle O_{\Delta}(t)O_{\Delta_3}(\omega_3)O_{\Delta_4}(\omega_4)\rangle.
\end{align}
This equals (We ignore overall numerical factors),
\small
\begin{align}
     W^{(s)}_{\Delta}(\omega_1,\omega_2,\omega_3,\omega_4)=\int_{-\infty}^{\infty}\cdots \int_{-\infty}^{\infty} \frac{dt dx dq_1 dq_2 dq_3}{q_1^{2\Delta-1}} e^{-iq_1(x-t)-iq_2 t-iq_3 x
     }\langle O_{\Delta_1}(\omega_1)O_{\Delta_2}(\omega_2)O_{\Delta}(q_3)\rangle\langle O_{\Delta}(q_3)O_{\Delta_3}(\omega_3)O_{\Delta_4}(\omega_4)\rangle.
\end{align}
\normalsize
Performing the integrals over $x$ and $t$, we obtain,
\small
\begin{align}\label{CPW1}
      W^{(s)}_{\Delta}(\omega_1,\omega_2,\omega_3,\omega_4)&=\int_{-\infty}^{\infty}\cdots \int_{-\infty}^{\infty} \frac{dq_1 dq_2 dq_3}{q_1^{2\Delta-1}}\delta(q_1+q_3)\delta(q_1-q_2)\langle O_{\Delta_1}(\omega_1)O_{\Delta_2}(\omega_2)O_{\Delta}(q_3)\rangle\langle O_{\Delta}(q_3)O_{\Delta_3}(\omega_3)O_{\Delta_4}(\omega_4)\rangle\notag\\
      &=\int_{-\infty}^{\infty} \frac{dq_1}{q_1^{2\Delta-1}}\langle O_{\Delta_1}(\omega_1)O_{\Delta_2}(\omega_2)O_{\Delta}(-q_1)\rangle\langle O_{\Delta}(q_1)O_{\Delta_3}(\omega_3)O_{\Delta_4}(\omega_4)\rangle\notag\\
      &=\int_{-\infty}^{\infty}\frac{dq_1}{q_1^{2\Delta-1}}\delta(\omega_1+\omega_2-q_1)\delta(q_1+\omega_3+\omega_4)\langle\langle O_{\Delta_1}(\omega_1)O_{\Delta_2}(\omega_2)O_{\Delta}(-q_1)\rangle\rangle\langle\langle O_{\Delta}(q_1)O_{\Delta_3}(\omega_3)O_{\Delta_4}(\omega_4)\rangle\rangle\notag\\
      &=\frac{\delta(\omega_1+\omega_2+\omega_3+\omega_4)}{(\omega_1+\omega_2)^{2\Delta-1}}\langle\langle O_{\Delta_1}(\omega_1)O_{\Delta_2}(\omega_2)O_{\Delta}(-\omega_1-\omega_2)\rangle\rangle\langle\langle O_{\Delta}(\omega_1+\omega_2)O_{\Delta_3}(\omega_3)O_{\Delta_4}(\omega_4)\rangle\rangle,
\end{align}
\normalsize
where we have defined the double bracket notation,
\begin{align}
    \langle .\rangle=\delta(\omega_1+\cdots)\langle\langle~.~\rangle\rangle.
\end{align}
Having obtained the conformal partial wave in momentum space \eqref{CPW1}, we can now extract the conformal block. We then use the expression for generic three point functions that we obtained in \eqref{3pointSol} in \eqref{CPW1}. Recall however, that we have two, rather than a unique solution to the three point function. As we explained in the beginning of this section and as we shall elaborate in subsection \ref{sec:TimeOrderingStuff}, this fact owes itself to the possibility of the various possible time orderings yielding different expressions when Fourier transformed. Let us now write \eqref{CPW1} explicitly. First, we define,
\begin{align}\label{fijk}
    f_{\Delta_i,\Delta_j,\Delta_k}(\omega_i,\omega_j)=\omega_i^{\Delta_i+\Delta_j+\Delta_k-2}&\bigg(c_{1,ijk}~_2 F_1\left(2-\Delta_i-\Delta_j-\Delta_k,1+\Delta_i-\Delta_j-\Delta_k,2(1-\Delta_j);\frac{-\omega_j}{\omega_i}\right)\notag\\
    &+\left(\frac{\omega_j}{\omega_i}\right)^{2\Delta_j-1}c_{2,ijk}~_2 F_1\left(1-\Delta_i+\Delta_j-\Delta_k,\Delta_i+\Delta_j-\Delta_k,2\Delta_j;-\frac{\omega_j}{\omega_i}\right)\bigg).
\end{align}
We can then re-write \eqref{CPW1} as,
\begin{align}\label{CPW2}
    W^{(s)}_{\Delta}(\omega_1,\omega_2,\omega_3,\omega_4)=\frac{1}{(\omega_1+\omega_2)^{2\Delta-1}}f_{\Delta_1,\Delta_2,\Delta}(\omega_1,\omega_2)f_{\Delta,\Delta_3,\Delta_4}(\omega_1+\omega_2,\omega_3)\delta(\omega_1+\omega_2+\omega_3+\omega_4),
\end{align}
which is our final expression for the $s$ channel momentum space conformal partial wave.  We will put \eqref{CPW2} to the test in subsection (\ref{sec:Examples}).

As we have seen throughout this subsection, we have found multiple solutions for the momentum space three and four point functions (\eqref{3pointSol} and \eqref{4pointSol}) and now for the conformal partial waves \eqref{CPW2}. Through an example, we shall now explain why there exist multiple solutions in momentum space in contrast to the unique expressions in the time domain.

\subsection{Time Ordering and the Existence of Multiple Solutions}\label{sec:TimeOrderingStuff}
To illustrate the existence of multiple momentum space solutions to the conformal Ward identities, let us compute a three point function via three different routes. First, we directly compute it in momentum space, then we Fourier transform the time domain correlators for the various possible time orderings, and finally we do so by using our solution \eqref{3pointSol} to the conformal ward identity.
\subsection*{Direct Momentum Space Computation}
Consider a three point function of a $\Delta=-1$ scalar operator. Its three point function can be expressed as the following one loop integral\footnote{This is the one dimensional cousin of the triple K integral \cite{Bzowski:2013sza}.}:
\begin{align}\label{OOOloopInt}
\nonumber\left<O(\om_1)O(\om_2)O(\om_3)\right>&=\delta(\om_1+\om_2+\om_3)\int_{-\infty}^{\infty} dl\,\frac{1}{l^2 (l-\om_1)^2 (l+\om_2)^2}\\
&=\delta(\om_1+\om_2+\om_3)\Tilde{G}(\omega_1,\omega_2).
\end{align}
We now employ an appropriate $i\epsilon$ prescription and use the residue theorem to evaluate \eqref{OOOloopInt}.
Enclosing all the poles, the result identically vanishes.
 Enclosing just one out of three poles yields three possible results:
\begin{equation}\label{Wick3point}
\langle O(\om_1)O(\om_2)O(\om_3)\rangle=4\pi i\delta(\omega_1+\omega_2+\omega_3)
\begin{cases}
\dfrac{2(\omega_1+2 \omega_2)}{\omega_2^3(\omega_1+\omega_2)^3},\, \text{Enclosing the pole at} \, -\omega_2. \\\\
-\dfrac{2 (\om_1-\om_2)}{\om_1^3 \om_2^3}, \,  \text{Enclosing the pole at} \, 0.  \\
\\
 - \dfrac{2 (2\om_1+\om_2)}{\om_1^3(\om_1+\om_2)^3},\,  \text{Enclosing the pole at} \, \omega_1.
\end{cases}
\end{equation}
The other three cases are the inclusion of two poles which give the same result (upto an overall negative sign) as the three above. 
\subsection*{Direct Fourier Transform}
We now directly Fourier transform the time domain $\langle OOO\rangle$ correlator. It is given by the following expression\footnote{This expression can be obtained by setting $\Delta_1=\Delta_2=\Delta_3=-1$ in \eqref{234pointtspace}.}:
\begin{align}\label{OOOtspace}
\left<O(t_1)O(t_2)O(t_3)\right>=\dfrac{C_{123}}{(t_1-t_2)^{-1}(t_1-t_3)^{-1}(t_2-t_3)^{-1}}=C_{123}(t_1-t_2)(t_1-t_3)(t_2-t_3).
\end{align}
Let us now Fourier transform \eqref{OOOtspace} for the various time orderings that are possible.
\subsubsection*{$\mathbf{t_1>t_2>t_3}$}
In this time ordering, the Fourier transform reads,
\begin{align}
  \nonumber \left<O(\om_1)O(\om_2)O(\om_3)\right>&=\int_{-\infty}^{\infty}dt_1\int_{-\infty}^{t_1} dt_2 \int_{-\infty}^{t_2}dt_3\, \left<O(t_1)O(t_2)O(t_3)\right> e^{i \om_1 t_1}e^{i \om_2 t_2}e^{i \om_3 t_3},\\
   &=\delta(\om_1+\om_2+\om_3) \,\dfrac{2 i (2\om_1+\om_2)}{\om_1^3(\om_1+\om_2)^3}.
\end{align}
\subsubsection*{$\mathbf{t_1>t_3>t_2}$}
Here we have,
\begin{align}
   \nonumber \left<O(\om_1)O(\om_2)O(\om_3)\right>&=\int_{-\infty}^{\infty}dt_1\int_{-\infty}^{t_1} dt_3 \int_{-\infty}^{t_3}dt_2\, \left<O(t_1)O(t_2)O(t_3)\right> e^{i \om_1 t_1}e^{i \om_2 t_2}e^{i \om_3 t_3},\\
   &=\delta(\om_1+\om_2+\om_3)\, \dfrac{2 i (\om_1-\om_2)}{\om_1^3 \om_2^3}.
\end{align}
Repeating this procedure for the remaining time orderings, we obtain their Fourier transforms as well. We summarize our results in table \ref{tab:my_label}.
\begin{table}[h]
 \renewcommand*{\arraystretch}{2.1}
    \centering
    \begin{tabular}{|l|c|}
    \hline
         \textbf{Time Ordering}& \textbf{Correlator}\\
         \hline
         $t_1>t_2>t_3$&  $\dfrac{2 i (2\om_1+\om_2)}{\om_1^3(\om_1+\om_2)^3}$\\
        \hline
         $t_2>t_1>t_3$&  $-\dfrac{2 i (\om_1+2\om_2)}{\om_2^3(\om_1+\om_2)^3}$\\
         \hline
         $t_3>t_2>t_1$&  $\dfrac{2 i (2\om_1+\om_2)}{\om_1^3(\om_1+\om_2)^3}$\\
         \hline
         $t_2>t_3>t_1$&  $\dfrac{2 i (\om_1-\om_2)}{\om_1^3\om_2^3}$\\
         \hline
         $t_1>t_3>t_2$&  $\dfrac{2 i (\om_1-\om_2)}{\om_1^3\om_2^3}$\\
         \hline
         $t_3>t_1>t_2$&  $-\dfrac{2 i (\om_1+2\om_2)}{\om_2^3(\om_1+\om_2)^3}$\\
         \hline
    \end{tabular}
    \caption{Correlators in momentum space obtained via Fourier Transform with different time orderings}
    \label{tab:my_label}
\end{table}

Notice that just like in \eqref{Wick3point}, we obtain three possible expressions for the momentum space correlator. However, note that only two of them are linearly independent. Therefore, the Fourier transform of every possible time ordering corresponds to a linear combination of these two expressions. Let us now see how the results of these various pole enclosures/ time orderings fit in with the general solution \eqref{3pointSol} that we obtained by solving the conformal Ward identities.
\subsection*{Comparison with the General Solution}
Our general solution for three point functions \eqref{3pointSol} for $\Delta_1=\Delta_2=\Delta_3=-1$ (recall that our operator $O$ has scaling dimension $-1$) is given by:
\begin{align}\label{Gtilde}
 \tilde{G}(\omega_1,\omega_2)&=\om_1^{-5}\left[c_1 \, \dfrac{\om_1^2(2\om_1+\om_2)}{2(\om_1+\om_2)^3}+c_2 \, \dfrac{\om_1^2(\om_1-\om_2)}{\om_2^3}\right]\\
     &=c_1 \, \dfrac{2\om_1+\om_2}{\om_1^3(\om_1+\om_2)^3}+c_2 \, \dfrac{(\om_1-\om_2)}{\om_1^3\om_2^3}.
\end{align}
The expression \eqref{Gtilde} covers all the cases we find in table \ref{tab:my_label} by appropriately choosing the constants $c_1$ and $c_2$.
The true meaning of the existence of the two independent solutions in \eqref{3pointSol} is now clear. They correspond to the fact that the different "time orderings" give rise to two possible Fourier space expressions (see table \ref{tab:my_label}). A similar analysis was performed in higher dimensions in \cite{Gillioz:2021sce}. The advantage of our analysis in $d=1$ is that the physical meaning of the existence of these multiple solutions is not obscured by technical complications.

Similarly, for the Four-Point function, we obtained four linearly independent solutions in equation \eqref{4pointSol}. One would suspect that these solutions should correspond to the Fourier transform of the various possible time orderings and is indeed what we see from a preliminary analysis. We also expect the same conclusions to hold for the $2^{n-2}$ independent solutions that we obtained for the $n$ point functions \eqref{nPointSol}.

We now proceed to verify our formulae for three, four point functions (\eqref{3pointSol}, \eqref{4pointSol}) and conformal partial waves \eqref{CPW2} by reproducing correlators in the free theories and the DFF model.
\subsection{Examples}\label{sec:Examples}
Let us begin by showing that three and four point correlators in the free bosonic theory are captured by our general formulae for correlation functions. 
\subsubsection{\normalsize{Free Bosonic Theory}}
The action for the $U(1)$ free bosonic theory is given by,
\begin{equation}
    S_{FB}=\int_{-\infty}^{\infty} dt~\partial_t\Bar{\phi}\partial_t\phi.
\end{equation}
$\phi$ and $\Bar{\phi}$ are primary operators with scaling dimension $-\frac{1}{2}$. We also consider the following composite primary operators:
\begin{align}
    &O(t)=\Bar{\phi}(t)\phi(t),~J_B(t)=i(\Bar{\phi}\partial_t\phi-\partial_t \Bar{\phi}~\phi),
\end{align}
which have the following momentum space counterparts:
\begin{align}\label{Opsomegaspace}
    &O(\omega)=\int_{-\infty}^{\infty} dl~\Bar{\phi}(l)\phi(\omega-l),~J_B(\omega)=\int_{-\infty}^{\infty} dl~(2l-\omega)\Bar{\phi}(l)\phi(\omega-l).
\end{align}
The first of these is a $\Delta=-1$ scalar while the second is the conserved $U(1)$ current. Let us now compute several three and four point correlators involving $\phi,\Bar{\phi}$ and these composite operators.
\subsubsection*{$\underline{\langle \Bar{\phi}(\omega_1)\phi(\omega_2)O(\omega_3)\rangle}$}
Performing the Wick contractions we obtain,
\begin{equation}\label{phibarphiO}
   \langle \Bar{\phi}(\omega_1)\phi(\omega_2)O(\omega_3)\rangle=\frac{2}{\omega_1^2\omega_2^2}\delta(\omega_1+\omega_2+\omega_3).
\end{equation}
We see that we can easily express \eqref{phibarphiO} in the form of our general solution \eqref{3pointSol}. Setting $c_1=0$ and $c_2=2$ in \eqref{3pointSol} we see that the result matches \eqref{phibarphiO}:
\begin{equation}
     \langle \Bar{\phi}(\omega_1)\phi(\omega_2)O(\omega_3)\rangle=\omega_1^{-4}\left(\frac{\omega_2}{\omega_1}\right)^{-2}~_2 F_1\left(2,0,-1,-\frac{\omega_2}{\omega_1}\right)\delta(\omega_1+\omega_2+\omega_3).
\end{equation}
Let us now consider a correlator involving the $U(1)$ current.
\subsubsection*{$\underline{\langle \Bar{\phi}(\omega_1)\phi(\omega_2)J_B(\omega_3)\rangle}$}
The Wick contractions yield,
\begin{equation}\label{phiphibarJ}
    \langle \Bar{\phi}(\omega_1)\phi(\omega_2)J_B(\omega_3)\rangle=\frac{\omega_1-\omega_2}{\omega_1^2\omega_2^2}\delta(\omega_1+\omega_2+\omega_3).
\end{equation}
In terms of the general solution \eqref{3pointSol} we see that,
\begin{equation}\label{phiphibarJsol}
     \langle \Bar{\phi}(\omega_1)\phi(\omega_2)J(\omega_3)\rangle=(\omega_1)^{-3}\left(\frac{\omega_2}{\omega_1}\right)^{-2}~_2 F_1\left(1,-1,-1,-\frac{\omega_2}{\omega_1}\right),
\end{equation}
providing another check of our result. Note that \eqref{phiphibarJ} also satisfies the charge conservation Ward Takahashi identity. We have,
\begin{align}
    \omega_3\langle\Bar{\phi}(\omega_1)\phi(\omega_2)J(\omega_3)\rangle=\left(\frac{1}{\omega_1^2}-\frac{1}{\omega_2^2}\right)\delta(\omega_1+\omega_2+\omega_3)=\big(\langle \phi(\omega_1)\Bar{\phi}(-\omega_1)\rangle-\langle \Bar{\phi}(\omega_2)\phi(-\omega_2)\rangle\big)\delta(\omega_1+\omega_2+\omega_3),
\end{align}
which shows that our general solution \eqref{phiphibarJsol} accommodates correlators involving conserved currents.

Let us now move on to a four point function.
\subsubsection*{$\underline{\langle \Bar{\phi}(\omega_1)O(\omega_2)O(\omega_3)\phi(\omega_4)\rangle}$}
Using the definitions of the operators provided in \eqref{Opsomegaspace} we obtain,
\begin{align}\label{phiOOphicorr}
     &\langle \phi(\omega_1) O(\omega_2) O(\omega_3)\Bar{\phi}(\omega_4)\rangle=\int_{-\infty}^{\infty}\int_{-\infty}^{\infty} dl dk\langle \phi(\omega_1)\Bar{\phi}(l)\phi(\omega_2-l)\Bar{\phi}(k)\phi(\omega_3-k)\Bar{\phi}(\omega_4)\rangle\notag\\&
    =\frac{1}{(2\pi)^2\omega_1^2\omega_4^2}\int_{-\infty}^{\infty}\int_{-\infty}^{\infty} dl dk\bigg[\frac{1}{k^2}\delta(\omega_1+l)\delta(\omega_2-l+k)\delta(\omega_3-k+\omega_4)+\frac{1}{l^2}\delta(l+\omega_3+k)\delta(\omega_2-l+\omega_4)\delta(\omega_1+k)\bigg]\notag\\
    &=\delta(\omega_1+\omega_2+\omega_3+\omega_4)\omega_1^{-6}\frac{1}{(1+x+y)^2}\bigg(\frac{1}{(1+x)^2}+\frac{1}{(1+y)^2}\bigg),
\end{align}
where $x=\frac{\omega_2}{\omega_1}$ and $y=\frac{\omega_3}{\omega_1}$ as we defined earlier.\\
We now define for convenience,
\begin{align}\label{phiOOphi}
    \langle \phi(\omega_1) O(\omega_2) O(\omega_3)\Bar{\phi}(\omega_4)\rangle=\delta(\omega_1+\omega_2+\omega_3+\omega_4)\omega_1^{-6}\psi_{\phi O O\Bar{\phi}}(x,y),
\end{align}
where,
\begin{align}
    \psi_{\phi O O\Bar{\phi}}(x,y)=\frac{1}{(1+x+y)^2}\bigg(\frac{1}{(1+x)^2}+\frac{1}{(1+y)^2}\bigg).
\end{align}
As we saw in \eqref{4pointSol}, the ward identities do not fix the form of the 4 point function and lead to an undetermined parameter $b_1$. Our aim now is to find value(s) of $b_1$ that reproduce this correlator.\\
We have $\Delta_1=\Delta_4=-\frac{1}{2}$ and $\Delta_2=\Delta_3=-1$. One of the solutions to the Ward identities \eqref{4pointSol} is,
\begin{align}
    \omega_1^{-6} f_{b_1}(x,y)=\omega_1^{-6}F_2(4,b_1,6-b_1,4,4;-x,-y).
\end{align}
Consider the following identity (see \cite{bateman1953higher}):
\begin{align}
F_2(a,b_1,b_2,a,a;-x,-y)=(1+x)^b_1(1+y)^{6-b_1}~_2 F_1\left(b_1,6-b_1,a,\frac{xy}{(1+x)(1+y)}\right).
\end{align}
Using this , we find that,
\begin{align}
 &F_2(4,2,4,4,4;-x,-y)=\frac{1}{(1+x+y)^2(1+y)^2},\notag\\
 &F_2(4,4,2,4,4;-x,-y)=\frac{1}{(1+x+y)^2(1+x)^2}.
\end{align}
which correspond to the choices $b_1=2$ and $b_2=4$ respectively. This gives us,
\begin{align}
    \omega_1^{-6}(f_2(x,y)+f_4(x,y))=\omega_1^{-6}\frac{1}{(1+x+y)^2}\bigg(\frac{1}{(1+x)^2}+\frac{1}{(1+y)^2}\bigg).
\end{align}
Comparing with \eqref{phiOOphi} we find,
\begin{align}
    \psi_{\phi O O \phi}=\bigg(F_2(4,2,4,4,4;-x,-y)+F_2(4,4,2,4,4;-x,-y)\bigg).
\end{align}
Therefore, the correlator \eqref{phiOOphicorr} can be written as,
\begin{align}
    \langle \phi(\omega_1) O(\omega_2) O(\omega_3)\Bar{\phi}(\omega_4)\rangle=\delta(\omega_1+\omega_2+\omega_3+\omega_4)\omega_1^{-6}\bigg(F_2\left(4,2,4,4,4;-\frac{\omega_2}{\omega_1},-\frac{\omega_3}{\omega_1}\right)+F_2\left(4,4,2,4,4;-\frac{\omega_2}{\omega_1},-\frac{\omega_3}{\omega_1}\right)\bigg),
\end{align}
which provides a check of our general result \eqref{4pointSol}. Let us now also reproduce this correlator using conformal partial waves. Consider the s channel conformal partial wave \eqref{CPW2}, set $c_{2,ijk}=0$ in \eqref{fijk} and $\Delta_1=\Delta_4=-\frac{1}{2},\Delta_2=\Delta_3=-1$. For the exchanged operator having $\Delta=-\frac{1}{2}$ we find,
\begin{align}
    W_{\Delta=-\frac{1}{2}}^{(s)}=\frac{1}{(1+x)^2(1+x+y)^2\omega_1^6
    }\delta(\omega_1+\omega_2+\omega_3+\omega_4).
\end{align}
We can also obtain the u channel conformal partial wave by a $(2\leftrightarrow 3)$ exchange and add it with the s channel result which yields,
\begin{align}
    W_{\Delta=-\frac{1}{2}}^{(s)}+W_{\Delta=-\frac{1}{2}}^{(u)}=\omega_1^{-6}\frac{1}{(1+x+y)^2}\bigg(\frac{1}{(1+x)^2}+\frac{1}{(1+y)^2}\bigg)\delta(\omega_1+\omega_2+\omega_3+\omega_4)=\langle \phi(\omega_1) O(\omega_2) O(\omega_3)\Bar{\phi}(\omega_4)\rangle.
\end{align}
To summarize, we have obtained two distinct representations for this correlator: One in terms of the momentum space conformal partial waves and the other, in terms of the Appell $F2$ function.
\begin{align}
    \langle \phi(\omega_1) O(\omega_2) O(\omega_3)\Bar{\phi}(\omega_4)\rangle&=W_{\Delta=-\frac{1}{2}}^{(s)}+W_{\Delta=-\frac{1}{2}}^{(u)}\notag\\&=\delta(\omega_1+\omega_2+\omega_3+\omega_4)\omega_1^{-6}\bigg(F_2
    \left(4,2,4,4,4;-\frac{\omega_2}{\omega_1},-\frac{\omega_3}{\omega_1}\right)+F_2\left(4,4,2,4,4;-\frac{\omega_2}{\omega_1},-\frac{\omega_3}{\omega_1}\right)\bigg).
\end{align}
Let us now provide two more examples of correlators in the free bosonic theory. Since the details of the calculation are similar to the above example, we just provide the final results.
\subsubsection*{$\underline{\langle O_1(\omega_1)O_1(\omega_2)O_1(\omega_3)O_1(\omega_4)\rangle}$}
The Fourier space expression for this correlator with the time ordering $t_1>t_2>t_3>t_4$ reads\footnote{One way in which this expression can be obtained is by performing the Wick contractions in the time domain and then Fourier transforming with this time ordering with the appropriate $i\epsilon$ prescriptions.},
\small
\begin{align}
    \langle\langle  O_1(\omega_1)O_1(\omega_2)O_1(\omega_3)O_1(\omega_4)\rangle\rangle=\frac{2(1+x)^2(41+7x(4+x))+2(1+x)(54+x(31+7x))y+4(10+x(5+x))y^2}{(1+x)^5(1+x+y)^3\omega_1^7}.
\end{align}
\normalsize
We find that this correlator receives contributions from \textit{three} different conformal blocks corresponding to exchanges of operators with $\Delta=-2,-1,$ and $0$. Setting $c_{2,ijk}=0$ and $\Delta_1=\Delta_2=\Delta_3=\Delta_4=-1$ in \eqref{CPW2}, we see that,
\begin{align}\label{OOOOCPW}
     \langle  O_1(\omega_1)O_1(\omega_2)O_1(\omega_3)O_1(\omega_4)\rangle=16 W_{\Delta=-1}^{(s)}-\frac{2}{3}W_{\Delta=0}^{(s)}+\frac{200}{3}W_{\Delta=-2}^{(s)}.
\end{align}
We also find that this result can be expressed in terms of our general four point function \eqref{4pointSol}. For $\Delta_1=\Delta_2=\Delta_3=\Delta_4=-1$ and setting $k_2=k_3=k_4=0$ in \eqref{4pointSol} we obtain the following two possible expressions (corresponding to the two choices of $a,b_1+b_2$ in \eqref{relabelling}):
\begin{align}\label{OOOO4pt2sols}
    f_{1}(b_1)=\frac{1}{\omega_1^7}F_2(7,b_1,4-b_1,4,4,-x,-y)~\text{and}~f_{2}(b_1)=\frac{1}{\omega_1^7}F_2(4,b_1,7-b_1,4,4,-x,-y).
\end{align}
It turns out that the second solution in \eqref{OOOO4pt2sols} is the required one for this example. Consider the following linear combination of the $b_1=4$, $b_1=5$ and $b_1=6$ solutions:
\begin{align}\label{OOOObeta}
     &\langle  O_1(\omega_1)O_1(\omega_2)O_1(\omega_3)O_1(\omega_4)\rangle=\delta(\omega_1+\omega_2+\omega_3+\omega_4)\big(14 f_2(4)+28 f_2(5)+40 f_2(6)\big)\notag\\
     &=\frac{\delta(\omega_1+\omega_2+\omega_3+\omega_4)}{\omega_1^7
     }\big(14 F_2(4,4,3,4,4,-x,-y)+28 F_2(4,5,2,4,4,-x,-y)+40 F_2(4,6,1,4,4,-x,-y)\big).
\end{align}
Let us now compare the conformal partial wave representation \eqref{OOOOCPW} and the Appell $F_2$ representation of the correlator \eqref{OOOObeta}. In \eqref{OOOOCPW}, it takes the sum of three different conformal blocks to reproduce the correlator. In \eqref{OOOObeta}, it takes three different ``$b_1$ exchanges" to reproduce the correlator. In lieu of this, one might think that $b_1$ is somehow related to the dimension of the exchanged operators. This, however, is not quite true as the number of conformal blocks does not always tally up with the number of ``$b_1$ exchanges".
\subsubsection*{$\underline{\langle J_B(\omega_1)J_B(\omega_2)O_1(\omega_3)O_1(\omega_4)\rangle}$}
The momentum space correlator corresponding to the time ordering $t_1>t_2>t_3>t_4$ is given by,
\begin{align}
    \langle\langle J_B(\omega_1)J_B(\omega_2)O_1(\omega_3)O(\omega_4)\rangle\rangle=\frac{1}{(1+x)(1+x+y)^3\omega_1^5}.
\end{align}
It turns out that a single conformal block suffices to reproduce this correlator. Setting $\Delta_1=\Delta_2=0,\Delta_3=\Delta_4=-1$ and $c_{2,ijk}=0$ in \eqref{CPW2} we see that,
\begin{align}
    \langle J_B(\omega_1)J_B(\omega_2)O_1(\omega_3)O(\omega_4)\rangle=W_{\Delta=0}^{(s)}.
\end{align}
In terms of our general four point solution \eqref{4pointSol}, set $\Delta_1=\Delta_2=0,\Delta_3=\Delta_4=-1$ and $k_2=k_3=k_4=0$ to obtain the two solutions:
\begin{align}
    f_{1}(b_1)=\frac{1}{\omega_1^5}F_2(5,b_1,4-b_1,2,4;-x,-y)~\text{and}~f_2(b_1)=\frac{1}{\omega_1^5}F_2(4,b_1,5-b_1,2,4;-x,-y).
\end{align}
We see that we can reproduce this correlator by a single ``$b_1$ exchange":
\begin{align}
     \langle J_B(\omega_1)J_B(\omega_2)O_1(\omega_3)O(\omega_4)\rangle=\delta(\omega_1+\omega_2+\omega_3+\omega_4)f_2(2)=\delta(\omega_1+\omega_2+\omega_3+\omega_4)F_2(4,2,3,2,4;-x,-y).
\end{align}
\subsubsection{\normalsize{Free Fermionic Theory}}
The action for the $U(1)$ free massless Dirac fermion theory is,
\begin{align}
    S_{FF}=i\int dt~\psi^\dagger \partial_t\psi.
\end{align}
$\psi$ and $\psi^\dagger$ are dimensionless operators. We also consider the conserved $U(1)$ current,
\begin{align}
    J_F(t)=\psi^\dagger(t)\psi(t),
\end{align}
which in momentum space is given by,
\begin{align}
    J_F(\omega)=\int dl~\psi^\dagger(l)\psi(\omega-l).
\end{align}
Let us now consider the following correlator:
\subsubsection*{$\underline{\langle \Bar{\psi}(\omega_1)\psi(\omega_2)J_F(\omega_3)\rangle}$}
Performing the Wick contractions gives,
\begin{equation}
    \langle \Bar{\psi}(\omega_1)\psi(\omega_2)O(\omega_3)\rangle=\frac{1}{\omega_1 \omega_2}\delta(\omega_1+\omega_2+\omega_3).
\end{equation}
This is reproduced by the general solution \eqref{3pointSol} in the following way:
\begin{equation}
    \langle\langle \Bar{\psi}(\omega_1)\psi(\omega_2)O(\omega_3)\rangle\rangle=(\omega_1)^{-2}\left(\frac{\omega_2}{\omega_1}\right)^{-1}~_2 F_1\left(1,0,0,\frac{-\omega_2}{\omega_1}\right).
\end{equation}
We now move on to a four-point example.
\subsubsection*{$\underline{\langle \psi^\dagger(\omega)J_F(\omega_2)J_F(\omega_3)\psi(\omega_4)\rangle}$}
Via Wick contractions, we obtain the following expression for this correlator:
\begin{align}
    \langle \psi^\dagger(\omega)J_F(\omega_2)J_F(\omega_3)\psi(\omega_4)\rangle=\frac{\delta(\omega_1+\omega_2+\omega_3+\omega_4)}{\omega_1^3}\frac{2+x+y}{(1+x)(1+y)(1+x+y)}.
\end{align}
This correlator is reproduced by the sum of a single conformal block in the s and u channels (obtained by setting $\Delta_1=\Delta_2=\Delta_3=\Delta_4=-1, c_{2,ijk}=0$ in \eqref{CPW2} and the u channel expression obtained via a $(2\leftrightarrow 3)$ exchange.):
\begin{align}
     \langle \psi^\dagger(\omega)J_F(\omega_2)J_F(\omega_3)\psi(\omega_4)\rangle=W_{\Delta=0}^{(s)}+W_{\Delta=0}^{(u)}.
\end{align}
As for the Appell function representation, set $\Delta_1=\Delta_2=\Delta_3=\Delta_4=0, k_2=k_3=k_4=0$ in \eqref{4pointSol}. We obtain,
\begin{align}
    f_1(b_1)=\frac{1}{\omega_1^3}F_2(3,b_1,2-b_1,2,2;-x,-y)~\text{and}~f_2(b_1)=F_2(2,b_1,3-b_1,2,2;-x,-y).
\end{align}
We find,
\begin{align}
    \langle \psi^\dagger(\omega)J_F(\omega_2)J_F(\omega_3)\psi(\omega_4)\rangle=\delta(\omega_1+\omega_2+\omega_3+\omega_4)f_1(1)=\delta(\omega_1+\omega_2+\omega_3+\omega_4)F_2(3,1,1,2,2;-x,-y),
\end{align}
providing another test of our results. Let us now consider and reproduce correlation functions in the DFF model.
\subsubsection{\normalsize{The DFF model}}
In this subsection, we use our general formulae for correlation functions to reproduce correlators in the DFF model. For details on the model, please refer to the original paper \cite{deAlfaro:1976vlx}. The four point function in the DFF model was computed in \cite{Jackiw:2012ur} and reads,
\begin{align}
    \langle O^\dagger_{r_0}(t_1)\phi_{\delta}(t_2)\phi_{\delta}(t_3)O_{r_0}(t_4)\rangle=\frac{1}{(t_{24}t_{13})^{\delta-r_0}(t_{12}t_{34})^{\delta+r_0}t_{14}^{2(r_0-\delta)}}\chi^{r_0}~_2 F_1(\delta,\delta,2r_0;\chi),
\end{align}
where we have used the shorthand $t_{ij}=t_i-t_j,$ $\displaystyle \chi=\frac{t_{12}t_{34}}{t_{13}t_{24}}$ is the cross ratio. The above expression is also in the time ordering $t_1>t_2>t_3>t_4$.\\\\ Consider the specific case where $r_0=\delta=-1$. The Fourier space expression of the correlator reads,
\begin{align}\label{DFFcorr1}
    \frac{(2+x)(2+2x+y)}{(1+x)^3(1+x+y)^3\omega_1^7}\delta(\omega_1+\omega_2+\omega_3+\omega_4).
\end{align}
This correlator (for arbitrary $\delta,r_0$) was found to receive contribution from just a single conformal block. Indeed, we see that just a single conformal block with $\Delta=-1$ exchange (obtained be setting $\Delta_1=\Delta_2=\Delta_3=\Delta_4=-1, \,c_{2,ijk}=0$ in \eqref{CPW2}) suffices to reproduce this correlator:
\begin{align}
     W_{\Delta=-1}^{(s)}=\frac{(2+x)(2+2x+y)}{(1+x)^3(1+x+y)^3\omega_1^7}\delta(\omega_1+\omega_2+\omega_3+\omega_4).
\end{align}
In terms of the Appell $F_2$ representation, set $\Delta_1=\Delta_2=\Delta_3=\Delta_4=-1,k_2=k_3=k_4=0$ in \eqref{4pointSol}. We obtain two solutions:
\begin{align}
    f_1(b_1)=\frac{1}{\omega_1^7}F_2(7,b_1,4-b_1,4,4;-x,-y)~\text{and}~f_2(b_1)=\frac{1}{\omega_1^7}F_2(4,b_1,7-b_1,4,4;-x,-y).
\end{align}
We see that,
\begin{align}
    \delta(\omega_1+\omega_2+\omega_3&+\omega_4)\frac{(2+x)(2+2x+y)}{(1+x)^3(1+x+y)^3\omega_1^7}=2\delta(\omega_1+\omega_2+\omega_3+\omega_4)\big(f_2(4)+f_2(5)\big)\notag\\
    &=\frac{2\delta(\omega_1+\omega_2+\omega_3+\omega_4)}{\omega_1^7}\big(F_2(4,4,3,4,4;-x,-y)+F_2(4,5,2,4,4;-x,-y)\big).
\end{align}
We can also repeat the same analysis for other values of $\delta$ and $r_0$. We find that the momentum space correlator for any $\delta,r_0$ is given in terms of exactly one conformal block in accordance with the results of \cite{Jackiw:2012ur}.

\section{Correlators in $\mathcal{N}=1$ Super Conformal Quantum Mechanics}\label{sec:Neq1SCQMcorrelators}
In this section, we shall extend our analysis to theories that have in addition to the $\mathfrak{sl}(2,\mathbb{R})$ conformal symmetry, $\mathcal{N}=1$ super symmetry. We first discuss the superspace formalism that we employ, which we then follow by solving the super conformal Ward identities for two, three and four point functions.
\subsection{The Superspace Formalism}
 The arena in which we work is the $\mathcal{N}=1$ superspace. A point in this superspace is described by the pair $(t,\theta)$ where $t$ is the usual time coordinate and $\theta$ is Grassmann valued. The generators of the $\mathcal{N}=1$ superconformal algebra consists of the usual conformal generators $H,K$ and $D$, the supersymmetry generator $Q$ and the special superconformal generator $S$. The algebra as well as the action of the generators on primary operators are given in appendix \ref{appendix:Neq1}.\\
In terms of component fields, the superfield can be expanded as follows:
\begin{equation}\label{susycomponentexpansion}
    \mathbf{O}_{\Delta}(\omega,\theta)=\Phi_{\Delta}(\omega)+\theta \,\Psi_{\Delta+\frac{1}{2}}(\omega).
\end{equation}
Our aim is to constrain the correlation functions of these superfields by solving the superconformal ward identities. These identities read,
\begin{align}\label{susyWardId}
\langle[\mathcal{L},\mathbf{O}_{\Delta_1}(\omega_1,\theta_1)]\cdots\mathbf{O}_{\Delta_n}(\omega_n,\theta_n)\rangle+\cdots \langle \mathbf{O}_{\Delta_1}(\omega_1,\theta_1)\cdots[\mathcal{L},\mathbf{O}_{\Delta_n}(\omega_n,\theta_n)]\rangle=0,~\mathcal{L}\in\{H,K,D,Q,S\}.
\end{align}
Using the action of the generators on primary operators provided in \eqref{Neq1Susygenerators} and \eqref{susyWardId} yields the following equations:
\begin{align}
\sum_{i=1}^n \omega_i f_n(\omega_1,\theta_1;\dots;\omega_n,\theta_n)&=0,\label{Neq1susyhWardId}\\
\sum_{i=1}^n\left(\omega_i\frac{\partial}{\partial \omega_i}+(1-\Delta_i)-\frac{1}{2}\theta_i \frac{\partial}{\partial \theta_i}\right) f_n(\omega_1,\theta_1;\dots;\omega_n,\theta_n)&=0,\label{Neq1susydWardId}\\
\sum_{i=1}^n\left(\omega_i \frac{\partial^2}{\partial\omega_i^2}+2(1-\Delta_i)\frac{\partial}{\partial \omega_i}-\theta_i\frac{\partial}{\partial \theta_i}\frac{\partial}{\partial \omega_i}\right) f_n(\omega_1,\theta_1;\dots;\omega_n,\theta_n)&=0,\label{Neq1susykWardId}\\
\sum_{i=1}^n \left(\frac{\partial}{\partial \theta_i}+\frac{\theta_i}{2}\omega_i\right) f_n(\omega_1,\theta_1;\dots;\omega_n,\theta_n)&=0,\label{Neq1susyqWardId}\\
\sum_{i=1}^n \left(\frac{\partial}{\partial \theta_i}\frac{\partial}{\partial \omega_i}+\left(\frac{1}{2}-\Delta_i\right)\theta_i+\frac{\theta_i \omega_i}{2}\frac{\partial}{\partial \omega_i}\right) f_n(\omega_1,\theta_1;\dots;\omega_n,\theta_n)&=0,\label{Neq1susysWardId}
\end{align}
where, $f_n(\omega_1,\theta_1;\dots;\omega_n,\theta_n)$ is a general $\mathcal{N}=1,$ $n$-point function.  Armed with the above equations, we now proceed to investigate its implications for correlation functions.
\subsection{Correlation Functions}
We begin with the two-point case.
\subsubsection{Two Point Functions}
Consider an arbitrary two-point function:
\begin{align}
    \langle \mathbf{O}_{\Delta_1}(\omega_1,\theta_1)\mathbf{O}_{\Delta_2}(\omega_2,\theta_2)\rangle=\langle \Phi_{\Delta_1}(\omega_1)\Phi_{\Delta_2}(\omega_2)\rangle-\theta_1\theta_2 \langle\Psi_{\Delta_1}(\omega_1)\Psi_{\Delta_2}(\omega_2)\rangle,
\end{align}
where we used the superfield expansion \footnote{Note that we did not retain any terms with an odd number of the $\theta_i$ as they multiply component correlators which are grassmann odd and hence zero.} \eqref{susycomponentexpansion}.\\
Translation, dilatation and special conformal invariance (equations \eqref{Neq1susyhWardId}, \eqref{Neq1susydWardId} and \eqref{Neq1susykWardId}) constrain the correlator to take the following form:
\begin{align}
    \langle \mathbf{O}_{\Delta_1}(\omega_1,\theta_1)\mathbf{O}_{\Delta_2}(\omega_2,\theta_2)\rangle=\delta(\omega_1+\omega_2)\delta_{\Delta_1,\Delta_2}\omega_1^{2\Delta_1-1}\big(c_0-c_1 \omega_1\theta_1 \theta_2\big).
\end{align}
The Q supersymmetric Ward identity \eqref{Neq1susyqWardId} then fixes $c_1=-\frac{c_0}{2}$. Therefore, the final result for the two point correlator reads,
\begin{align}\label{superNeq1twopoint}
    \langle \mathbf{O}_{\Delta_1}(\omega_1,\theta_1)\mathbf{O}_{\Delta_2}(\omega_2,\theta_2)\rangle=c_0\delta(\omega_1+\omega_2)\delta_{\Delta_1,\Delta_2}\omega_1^{2\Delta_1-1}\left(1+\frac{\omega_1}{2} \theta_1 \theta_2\right).
\end{align}
Note that the result is quite reminiscent of what was obtained in three dimensions in \cite{Jain:2023idr}.
\subsubsection{Three Point Functions}
We now move to the three point case. A generic correlator reads:
\begin{align}
   \langle \textbf{O}_{\Delta_1}(\omega_1,\theta_1)\textbf{O}_{\Delta_2}(\omega_2,\theta_2)\textbf{O}_{\Delta_3}(\omega_3,\theta_3)\rangle&=\delta(\omega_1+\omega_2+\omega_3)\big(c_0(\omega_1,\oemga_2)+c_4(\omega_1,\oemga_2)\theta_1\theta_2\notag\\
&\qquad+c_5(\omega_1,\oemga_2)\theta_2\theta_3+c_6(\omega_1,\oemga_2)\theta_1\theta_3\big).
\end{align}
This correlator is constrained by the $Q$ Ward identity \eqref{Neq1susyqWardId} which yields the following constraints:
\begin{equation}\label{c5c6constraint}
c_5(\omega_1,\omega_2)=\left(c_4(\omega_1,\omega_2)+\frac{1}{2}c_0(\omega_1,\omega_2)\oemga_2\right),\quad c_6(\omega_1,\omega_2)=\left(-c_4(\omega_1,\omega_2)+\frac{1}{2}c_0(\omega_1,\omega_2) \oemga_1\right).
\end{equation}
Our correlator thus takes the form:
\begin{align}
\langle \textbf{O}_{\Delta_1}(\omega_1,\theta_1)&\textbf{O}_{\Delta_2}(\omega_2,\theta_2)\textbf{O}_{\Delta_3}(\omega_3,\theta_3)\rangle=\delta(\omega_1+\omega_2+\omega_3)\Bigg(c_0(\omega_1,\omega_2)+c_4(\omega_1,\omega_2) \theta_1\theta_2\notag\\
&+\left(-c_4(\omega_1,\omega_2)+\frac{1}{2}c_0(\omega_1,\omega_2) \oemga_1\right)\theta_1\theta_3+\left(c_4(\omega_1,\omega_2)+\frac{1}{2}c_0(\omega_1,\omega_2) \oemga_2\right)\theta_2\theta_3\Bigg).
\end{align}
Dilatation invariance \eqref{Neq1susydWardId} then fixes the overall scaling of the correlator thus leading to,
\begin{align}
\langle \textbf{O}_{\Delta_1}(\omega_1,\theta_1)\textbf{O}_{\Delta_2}(\omega_2,\theta_2)\textbf{O}_{\Delta_3}(\omega_3,\theta_3)\rangle&=\delta(\omega_1+\omega_2+\omega_3)\omega_1^{\Delta_t-2}\bigg(c_0(x)\left(2+\oemga_1\,\theta_1\theta_3+\omega_1\, x\,\theta_2\theta_3\right)\notag\\
&\qquad\qquad+2 \omega_1\,c_4(x)(\theta_1\theta_2-\theta_1\theta_3+\theta_2\theta_3) \bigg),
\end{align}
where $\Delta_t=\Delta_1+\Delta_2+\Delta_3,~~ \displaystyle x=\frac{\omega_2}{\omega_1}$. 

The final step is to obtain constraints from the special conformal Ward identity \eqref{Neq1susykWardId} (The $S$ Ward identity will then trivially follows as $[K,Q]=-S$). This results in four differential equations out of which only three are independent, viz,
\begin{equation}
    x(1+x) \frac{d^2 c_0(x)}{d x^2}+2\Big(1-\Delta_2-x(-2+\Delta_2+\Delta_3)\Big)\frac{dc_0(x)}{dx}-(1+2\Delta_1-\Delta_t)(-2+\Delta_t)c_0(x)=0,\notag
    \end{equation}
    \begin{equation}
     x(1+x) \frac{d^2 c_4(x)}{d x^2}+\Big(1-2\Delta_2+x(3-2\Delta_2-2\Delta_3)\Big)\frac{dc_4(x)}{dx}-(1+2\Delta_1-\Delta_t)(-1+\Delta_t)c_4(x)=0,\notag
     \end{equation}
     \begin{align}
     x^2(1+x)\frac{d^2c_0(x)}{dx^2} +x\Big(3-2\Delta_2-2x(-2+\Delta_2+\Delta_3)\Big)\frac{d c_0(x)}{dx}&+\Big(1-2\Delta_2-x(1+2\Delta_1-\Delta_t)(-2+\Delta_t)\Big)c_0(x),\notag\\
    + 2x(1+x)\frac{d^2 c_4(x)}{dx^2}+\Big(2-4\Delta_2-4x(-1+\Delta_2+\Delta_3)\Big)\frac{d c_4(x)}{dx}&-2(2\Delta_1-\Delta_t)(-1+\Delta_t)c_4(x)=0.\notag
 \end{align}
 Solving the first equation gives the solution for $c_0(x)$
 \begin{align}
     c_0(x)&=c_{01}\, _2F_1\Big(2-\Delta_t,1+2\Delta_1-\Delta_t,2-2\Delta_2;-x\Big)\notag\\
     &\qquad \qquad+c_{02}\,x^{2\Delta_2-1}\, _2F_1\Big(1+2\Delta_2-\Delta_t,\Delta_t-2\Delta_3,2\Delta_2;-x\Big),
 \end{align}
 which is exactly the non supersymmetric correlator \eqref{3pointSol} as expected.
 Similarly, solving the second equation gives,
 \begin{align}
      c_4(x)&=c_{41}\, _2F_1\Big(1-\Delta_t,1+2\Delta_1-\Delta_t,1-2\Delta_2;-x\Big)\notag\\
     &\qquad \qquad+c_{42}\,x^{2\Delta_2}\, _2F_1\Big(1+2\Delta_2-\Delta_t,1+\Delta_t-2\Delta_3,1+2\Delta_2;-x\Big).
 \end{align}
 The third equation mixes the coefficients giving the following constraints,
 \begin{equation}
     c_{41}=c_{01}\frac{2\Delta_2-1}{2(-1+\Delta_t)},\qquad c_{42}=-c_{02}\frac{\Delta_t-2\Delta_3}{4\Delta_2}.
 \end{equation}
 Using the constraints obtained by $Q$ action \eqref{c5c6constraint}, and Hypergeometric function Identities given in appendix \ref{appendix:hypergeometric},
  we obtain the following form for $c_5(x)$:
 \begin{align}
c_5(x)&=c_{51}\, _2F_1\Big(1-\Delta_t,2\Delta_1-\Delta_t,1-2\Delta_2;-x\Big)\notag\\
     &\qquad \qquad+c_{52}\,x^{2\Delta_2}\, _2F_1\Big(1+2\Delta_2-\Delta_t,2\Delta_1+2\Delta_2-\Delta_t,1+2\Delta_2;-x\Big),
 \end{align}
 where the coefficients are given by, 
 \begin{align}
c_{51}&=c_{01}\frac{2\Delta_2-1}{2(-1+\Delta_t)},\quad c_{52}=c_{02}\frac{\Delta_t-2\Delta_1}{4\Delta_2}.
\end{align}
Similarly, we can repeat the same procedure to obtain $c_6(x)$:
 \begin{align}
c_6(x)&=c_{61}\, _2F_1\Big(1-\Delta_t,1+2\Delta_1-\Delta_t,2-2\Delta_2;-x\Big)\notag\\
     &\qquad \qquad+c_{62}\,x^{2\Delta_2-1}\, _2F_1\Big(2\Delta_2-\Delta_t,2\Delta_1+2\Delta_2-\Delta_t,2\Delta_2;-x\Big),
 \end{align}
with coefficients $c_{61}$ and $c_{62}$ given by,
\begin{align}c_{61}&=c_{01}\frac{\Delta_t-2\Delta_2}{2(-1+\Delta_t)},\quad c_{62}=\frac{c_{02}}{2}.
 \end{align}
 Therefore, our final expression for the three-point function in $\mathcal{N}=1$ SCQM reads,
 \begin{align}\label{Neq13pointSol}
     \langle \textbf{O}_{\Delta_1}(\omega_1,\theta_1)&\textbf{O}_{\Delta_2}(\omega_2,\theta_2)\textbf{O}_{\Delta_3}(\omega_3,\theta_3)\rangle=\delta(\omega_1+\omega_2+\omega_3)\omega_1^{\Delta_t-2}\notag\\
     &\Bigg(c_{01}\Big( \,_2F_1\left(2-\Delta_t,1+2\Delta_1-\Delta_t,2-2\Delta_2;-x\right)\left(2+\oemga_1\,\theta_1\theta_3+\omega_1\, x\,\theta_2\theta_3\right)\notag\\
    & \quad\quad+\omega_1\frac{2\Delta_2-1}{\Delta_t-1}\, _2F_1\left(1-\Delta_t,1+2\Delta_1-\Delta_t,1-2\Delta_2;-x\right)(\theta_1\theta_2-\theta_1\theta_3+\theta_2\theta_3)\Big)\notag\\
   & +c_{02}\,x^{2\Delta_2-1}\Big(\, _2F_1\left(1+2\Delta_2-\Delta_t,\Delta_t-2\Delta_3,2\Delta_2;-x\right)\left(2+\oemga_1\,\theta_1\theta_3+\omega_1\, x\,\theta_2\theta_3\right)\notag\\
  & \quad\quad-\omega_1 \,x\frac{\Delta_t-2\Delta_3}{2\Delta_2}\, _2F_1\left(1+2\Delta_2-\Delta_t,1+\Delta_t-2\Delta_3,1+2\Delta_2;-x\right)(\theta_1\theta_2-\theta_1\theta_3+\theta_2\theta_3)\Big)\Bigg).
 \end{align}
 where $\displaystyle x=\frac{\omega_2}{\omega_1}$.
\subsubsection{Four Point Functions}
We now move to the four-point case. A generic correlator reads:
\begin{align}
   \langle \textbf{O}_{\Delta_1}(\omega_1,\theta_1)\textbf{O}_{\Delta_2}(\omega_2,\theta_2)\textbf{O}_{\Delta_3}(\omega_3,\theta_3)\textbf{O}_{\Delta_4}(\omega_4,\theta_4)\rangle&=\delta(\omega_1+\omega_2+\omega_3+\omega_4)\\
&\big(c_0(\omega_1,\oemga_2,\omega_3)+c_{12}(\omega_1,\oemga_2,\omega_3)\theta_1\theta_2\notag\\
&+c_{13}(\omega_1,\oemga_2,\omega_3)\theta_1\theta_3+c_{14}(\omega_1,\oemga_2,\omega_3)\theta_1\theta_4\notag\\
&+c_{23}(\omega_1,\oemga_2,\omega_3)\theta_2\theta_3+c_{24}(\omega_1,\oemga_2,\omega_3)\theta_2\theta_4\notag\\
&+c_{34}(\omega_1,\oemga_2,\omega_3)\theta_3\theta_4+c_{1234}(\omega_1,\oemga_2,\omega_3)\theta_1\theta_2\theta_3\theta_4\big).
\end{align}
Constraining the correlator by the $Q$ Ward Identity \eqref{Neq1susyqWardId}, yields the following constraints:
\begin{equation}
    \begin{split}
 c_{1234}&=\frac{1}{2}\left(c_{23}\omega_1-c_{13}\omega_2+c_{12}\omega_3\right),\\
 c_{14}&=-c_{12}-c_{13}+\frac{c_0 \omega_1}{2},
    \end{split}
    \quad
    \begin{split}
       c_{24}&=c_{12}-c_{23}+\frac{c_0\omega_2}{2},\\
       c_{34}&=c_{13}+c_{23}+\frac{c_0\omega_3}{2}.
    \end{split}
\end{equation}
Dilatation invariance \eqref{Neq1susydWardId} then fixes the overall scaling of the correlator, thus leading to,
\begin{align}
\langle \textbf{O}_{\Delta_1}(\omega_1,\theta_1)\textbf{O}_{\Delta_2}(\omega_2,\theta_2)\textbf{O}_{\Delta_3}(\omega_3,\theta_3)\textbf{O}_{\Delta_4}(\omega_4,\theta_4)\rangle&=\delta(\omega_1+\omega_2+\omega_3+\omega_4)\omega_1^{\Delta_t-3}\\
&\bigg(c_0(x,y)\left(2+\oemga_1\,\theta_1\theta_4+\omega_2\,\,\theta_2\theta_4+\omega_3\,\theta_3\theta_4\right)\notag\\
&+\omega_1\,c_{23}(x,y)\big(2(\theta_2\theta_3-\theta_2\theta_4+\theta_3\theta_4)+\omega_1\theta_1\theta_2\theta_3\theta_4\big)\notag\\
&+\omega_1\,c_{13}(x,y)\big(2(\theta_1\theta_3-\theta_1\theta_4+\theta_3\theta_4)-\omega_2\theta_1\theta_2\theta_3\theta_4\big)\notag\\
&+\omega_1\,c_{12}(x,y)\big(2(\theta_1\theta_2-\theta_1\theta_4+\theta_2\theta_4)+\omega_3\theta_1\theta_2\theta_3\theta_4\big)
\bigg),
\end{align}
where $\Delta_t=\Delta_1+\Delta_2+\Delta_3+\Delta_4, \displaystyle x=\frac{\omega_2}{\omega_1},y=\frac{\omega_3}{\omega_1}$.

From the component expansion, it can be seen that the components $c_{12},\dots$ have the same functional form as $c_0(x,y)$, i.e., Appell functions \eqref{4pointSol}, but differ only by the scaling dimensions. However, we can follow the same routine as we followed for the three point function and apply $K$ to get the constraints on coefficients. We choose not to do that and instead apply $S$ as it has a first-order action and leads to simpler equations. If we fix the component correlators by their Appell function representation \eqref{4pointSol} and leave the coefficients undetermined, then apply $S$, we obtain constraints connecting these coefficients. For instance at $\order{\theta_1}$ and $\order{\theta_2}$ we find,
\begin{align}
    2 \frac{d c_{12}}{dx}+2\frac{d c_{13}}{dy}-x \frac{d c_0}{dx}-y \frac{d c_0}{dy}-(2+\Delta_1-\Delta_2-\Delta_3-\Delta_4)c_0&=0,\\
    -2x \frac{d c_{12}}{dx}-2y \frac{d c_{12}}{dy}+2\frac{d c_{23}}{dy}+x \frac{d c_0}{dx}-2(2-\Delta_t)c_{12}+(1-2\Delta_2)c_0&=0.
\end{align}
Proceeding to higher orders in the Grassmann expansion, we obtain similar constraints. To solve these equations (as well as the analysis for higher point functions) is an interesting problem which we defer to the future. We now proceed to the $\mathcal{N}=2$ case.
\section{Correlators in $\mathcal{N}=2$ Super Conformal Quantum Mechanics}\label{sec:Neq2SCQMcorrelators}
In this section, we extend our results to conformal theories that also possess $\mathcal{N}=2$ supersymmetry. Similar to section \ref{sec:Neq1SCQMcorrelators} we first introduce the superspace formalism in which we work and then solve the super conformal Ward identities for two and three point functions.
\subsection{The Superspace Formalism}
 The arena in which we work is the $\mathcal{N}=2$ superspace. A point in this superspace is described by the triplet $(t,\theta,\Bar{\theta})$ where $t$ is the usual time coordinate and $\theta$ is a complex Grassmann variable. The momentum (or rather, frequency) superspace is spanned by the triplet $(\omega,\theta,\Bar{\theta})$. The generators of the $\mathcal{N}=2$ superconformal algebra consists of the usual conformal generators $H,K$ and $D$, the supersymmetry generators $Q,\Bar{Q}$, the special superconformal generators $S,\Bar{S}$ and the $U(1)$, $R$-symmetry generator $R$. Their algebra can be found for instance in \cite{Okazaki:2015pfa}.
The action of all generators on primary (bosonic) superfields $\mathbf{O}_{\Delta}(\omega,\theta)$ are given in appendix \ref{appendix:Neq2}. We bring the reader's attention to the $R$ symmetry generator (last equation of \eqref{WardIdNeq2}). The  $\theta_i$ have $R$ charge $+1$ while the $\Bar{\theta}_i$ have $R$ charge $-1$. Therefore, every single term in a correlation function should contain an equal number of $\theta_i$ and $\Bar{\theta}_i$.

In the $\mathcal{N}=2$ superspace formalism, (bosonic) superfields can be expanded as follows:
\begin{equation}\label{Neq2susycomponentexpansion}
    \mathbf{O}_{\Delta}(\omega,\theta)=\Phi_{\Delta}(\omega)+\theta \,\Psi_{\Delta+\frac{1}{2}}(\omega)+\Bar{\theta}\,\Bar{\Psi}_{\Delta+\frac{1}{2}}(\omega)+\theta\Bar{\theta}F_{\Delta+1}(\omega).
\end{equation}
Our aim in what is to follow is to constrain the correlation functions of these superfields by solving the superconformal ward identities. These identities read,
\begin{align}\label{Neq2susyWardId}
    &\langle[\mathcal{L},\mathbf{O}_{\Delta_1}(\omega_1,\theta_1)]\cdots\mathbf{O}_{\Delta_n}(\omega_n,\theta_n)\rangle+\cdots \langle \mathbf{O}_{\Delta_1}(\omega_1,\theta_1)\cdots[\mathcal{L},\mathbf{O}_{\Delta_n}(\omega_n,\theta_n)]\rangle=0,~\mathcal{L}\in\{H,K,D,Q,S,\Bar{Q},\Bar{S},R\}.
\end{align}
The explicit form of these identities can be found in appendix \ref{appendix:Neq2}. Armed with the above equations, we now proceed to investigate its implications for correlation functions.
\subsection{Correlation functions}
In this subsection, we obtain two and three-point functions that were obtained by solving the $\mathcal{N}=2$ superconformal Ward identities. Since the procedure of obtaining the solutions is identical to that of the $\mathcal{N}=1$ case, we shall desist from providing details and instead provide the final results.
\subsubsection{Two Point Functions}
The two point function of a generic $\mathcal{N}=2$ primary superfield takes the following form:
\begin{align}\label{superNeq2twopoint}
    \langle \mathbf{O}_{\Delta_1}(\omega_1,\theta_1,\Bar{\theta}_1)\mathbf{O}_{\Delta_2}(\omega_2,\theta_2,\Bar{\theta}_2)\rangle=\delta_{\Delta_1,\Delta_2}\delta(\omega_1+\omega_2)\omega_1^{2\Delta_1-1}\bigg(4+2\omega_1(\theta_1\Bar{\theta}_2-\theta_2\Bar{\theta}_1)-\omega_1^2\theta_1\theta_2\Bar{\theta}_1\Bar{\theta}_2\bigg).
\end{align}
It can easily be verified that this expression satisfies the superconformal Ward identities. We also note an interesting relation between \eqref{superNeq2twopoint} and it's $\mathcal{N}=1$ counterpart \eqref{superNeq1twopoint}:
\begin{align}\label{DoubleCopy}
     \langle \mathbf{O}_{2\Delta_1}(\omega_1,\theta_1,\Bar{\theta}_1)\mathbf{O}_{2\Delta_2}(\omega_2,\theta_2,\Bar{\theta}_2)\rangle_{\mathcal{N}=2}=\omega_1\langle \mathbf{O}_{\Delta_1}(\omega_1,\theta_1)\mathbf{O}_{\Delta_2}(\omega_2,\Bar{\theta}_2)\rangle_{\mathcal{N}=1}~\langle \mathbf{O}_{\Delta_1}(\omega_1,\Bar{\theta}_1)\mathbf{O}_{\Delta_2}(\omega_2,\theta_2)\rangle_{\mathcal{N}=1}.
\end{align}
The above form is the unique product of $\mathcal{N}=1$ two point functions that possess the required $R$ symmetry. Indeed, this is reminiscent of the super double copy obtained in three-dimensional conformal field theories in \cite{Jain:2023idr}. The authors of \cite{Jain:2023idr} also obtained a super double copy at the three point level. Let us also thus move on to the three point case.
\subsubsection{Three Point Functions}
Let us first define,
\begin{align}
    C_{i}(\Delta_1,\Delta_2,\Delta_3;-x)&=c_{i1}~_2F_1(2-\Delta_1-\Delta_2-\Delta_3,1+\Delta_1-\Delta_2-\Delta_3,2(1-\Delta_2);-x)\notag\\
    &\quad\quad+x^{2\Delta_2-1}~c_{i2}~_2F_1(1-\Delta_1+\Delta_2-\Delta_3,\Delta_1+\Delta_2-\Delta_3,2\Delta_2,-x),
\end{align}
and,
\begin{align}
    c_0(x)=C_0
    \left(\Delta_1,\Delta_2,\Delta_3;-x\right),\quad& c_6(x)=C_6\left(\Delta_1+\frac{1}{2},\Delta_2+\frac{1}{2},\Delta_3;-x\right),\notag\\
    c_8(x)=C_8\left(\Delta_1+\frac{1}{2},\Delta_2,\Delta_3+\frac{1}{2};-x\right),\quad &c_9(x)=C_9\left(\Delta_1,\Delta_2+\frac{1}{2},\Delta_3+\frac{1}{2};-x\right),\notag\\
    c_{11}(x)=C_{11}\left(\Delta_1,\Delta_2+1,\Delta_3;-x\right),\quad & c_{12}(x)=C_{12}\left(\Delta_1+1,\Delta_2+1,\Delta_3;-x\right).
\end{align}
Our result for the $\mathcal{N}=2$ SCQM three-point correlator after solving the superconformal Ward identities \eqref{WardIdNeq2} is the following expression:
\begin{align}\label{Neq2threepoint}
    \langle \mathbf{O}_{\Delta_1}(\omega_1,\theta_1,\Bar{\theta}_1)&\mathbf{O}_{\Delta_2}(\omega_2,\theta_2,\Bar{\theta}_2)\mathbf{O}_{\Delta_3}(\omega_3,\theta_3,\Bar{\theta}_3)\rangle=\delta(\omega_1+\omega_2+\omega_3)\omega_1^{\Delta_t-2}\notag\\
    & \Bigg[c_0(x)\bigg(1+\frac{\omega_1}{4}\Big(-2\theta_1\bar\theta_1-2x \theta_1\bar\theta_2+2(x+2)\theta_1\bar\theta_3+2x \theta_2\bar\theta_3-2(1+x)\theta_3\bar\theta_3\notag\\
    &\qquad\qquad\qquad +x\, \omega_1 \theta_1\theta_2\bar\theta_1\bar\theta_3+x^2 \omega_1\theta_1\theta_2\bar\theta_2\bar\theta_3-\omega_1(1+x)\theta_1\theta_3\bar\theta_1\bar\theta_3-x(1+x)\omega_1\theta_1\theta_3\bar\theta_2\bar\theta_3\Big)\bigg)\notag\\ &+c_6(x)\bigg(\frac{\omega_1}{2}\Big(-2\theta_1\Bar{\theta}_1+2\theta_1\Bar{\theta}_3+2\theta_2\bar\theta_1-2\theta_2\bar\theta_3+\omega_1\theta_1\theta_2\bar\theta_1\bar\theta_3+x\omega_1\theta_1 \theta_2\bar\theta_1\bar\theta_3\notag\\
    &\qquad\qquad\qquad-x \omega_1\theta_1\theta_3\bar\theta_1\bar\theta_2-\omega_1\theta_1\theta_3\bar\theta_1\bar\theta_3-x\omega_1\theta_1\theta_3\bar\theta_2\bar\theta_3+x\omega_1\theta_2\theta_3\bar\theta_1\bar\theta_2\notag\\
    &\qquad\qquad\qquad+\omega_1\theta_2\theta_3\bar\theta_1\bar\theta_3+x \omega_1 \theta_2\theta_3\bar\theta_2\bar\theta_3\Big)+\frac{x(x+1)}{4}\omega_1^3\theta_1\theta_2\theta_3\bar\theta_1\bar\theta_2\bar\theta_3\bigg)\notag\\
    &+c_8(x)\bigg(\frac{\omega_1}{2}\Big(-2\theta_1\bar\theta_1+2\theta_1\bar\theta_3+2\theta_3\bar\theta_1-2\theta_3\bar\theta_3+x\omega_1\theta_1\theta_2\bar\theta_1\bar\theta_3-x\omega_1\theta_1\theta_3\bar\theta_1\bar\theta_2-x\omega_1\theta_1\theta_3\bar\theta_2\bar\theta_3\notag\\
    &\qquad\qquad\qquad+x\omega_1\theta_2\theta_3\bar\theta_1\bar\theta_3\Big)+\frac{x^2\omega_1^3}{4}\theta_1\theta_2\theta_3\bar\theta_1\bar\theta_2\bar\theta_3\bigg)\notag\\
    &+c_9(x)\bigg(\frac{\omega_1}{2}\Big(-2\theta_1\bar\theta_2+2\theta_1\bar\theta_3+2\theta_3\bar\theta_2-2\theta_3\bar\theta_3+x\omega_1\theta_1\theta_2\bar\theta_2\bar\theta_3+\omega_1\theta_1\theta_3\bar\theta_1\bar\theta_2-\omega_1\theta_1\theta_3\bar\theta_1\bar\theta_3\notag\\
    &\qquad\qquad\qquad+\omega_1\theta_1\theta_3\bar\theta_2\bar\theta_3-x\omega_1\theta_1\theta_3\bar\theta_2\bar\theta_3+x\omega_1\theta_2\theta_3\bar\theta_2\bar\theta_3\Big)-\frac{x}{4}\omega_1^3\theta_1\theta_2\theta_3\bar\theta_1\bar\theta_2\bar\theta_3\bigg)\notag\\
    &+c_{11}(x)\bigg(\frac{\omega_1}{2}\Big(-2\theta_1\bar\theta_2+2\theta_1\bar\theta_3+2\theta_2\bar\theta_2-2\theta_2\bar\theta_3+\omega_1\theta_1\theta_2\bar\theta_2\bar\theta_3+x\omega_1\theta_1\theta_2\bar\theta_2\bar\theta_3\notag\\
    &\qquad\qquad\qquad+\omega_1\theta_1\theta_3\bar\theta_1\bar\theta_2-\omega_1\theta_1\theta_3\bar\theta_1\bar\theta_3-x\omega_1\theta_1\theta_3\bar\theta_2\bar\theta_3-\omega_1\theta_2\theta_3\bar\theta_1\bar\theta_2\notag\\
    &\qquad\qquad\qquad+\omega_1\theta_2\theta_3\bar\theta_1\bar\theta_3+x\omega_1\theta_2\theta_3\bar\theta_2\bar\theta_3\Big)-\frac{(x+1)}{4}\omega_1^3\theta_1\theta_2\theta_3\bar\theta_1\bar\theta_2\bar\theta_3\bigg)\notag\\
    &+c_{12}(x)\omega_1^2\bigg((\theta_1\theta_2\bar\theta_1\bar\theta_2-\theta_1\theta_2\bar\theta_1\bar\theta_3+\theta_1\theta_2\bar\theta_2\bar\theta_3-\theta_1\theta_3\bar\theta_1\bar\theta_2+\theta_1\theta_3\bar\theta_1\bar\theta_3-\theta_1\theta_3\bar\theta_2\bar\theta_3+\theta_2\theta_3\bar\theta_1\bar\theta_2\notag\\
    &\qquad\qquad\qquad-\theta_2\theta_3\bar\theta_1\bar\theta_3)+\omega_1^2\theta_2\theta_3\bar\theta_2\bar\theta_3\bigg)\Bigg].
\end{align}
where $\displaystyle x=\frac{\omega_2}{\omega_1}$ and the coefficients of the $c_i$ are related as follows:
\begin{align}
\begin{split}
    c_{81}&=-c_{01}\frac{\Delta_1}{\Delta_t-1}-c_{61}
    \frac{\Delta_1+\Delta_2-\Delta_3}{2\Delta_2-1},
    \\
    c_{91}&=-\frac{c_{01}(2\Delta_2-1)(\Delta_1-\Delta_3)}{(2\Delta_1-\Delta_t)(\Delta_t-1)}+c_{61}\frac{(
    2\Delta_3-\Delta_t)
    }{(2\Delta_1-\Delta_t)},
    \\
    c_{121}&=c_{01}\frac{\Delta_2(1-2\Delta_2)}{2(\Delta_3-1)\Delta_t},\\
    c_{111}&=\Delta_2\frac{c_{01}(2\Delta_2-1)+2c_{61}(\Delta_t-1)}{(2\Delta_1-\Delta_t)(\Delta_t-1)},
    \end{split}
    \begin{split}
        c_{82}&=-\frac{c_{02}\Delta_1+2c_{62}\Delta_2}{\Delta_1-\Delta_2+\Delta_3},\\
        c_{92}&=\frac{c_{02}(\Delta_1-\Delta_3)-2c_{62}\Delta_2}{2\Delta_2},\\
        c_{122}&=-c_{02}\frac{(\Delta_1+\Delta_2-\Delta_3)(1+\Delta_1+\Delta_2-\Delta_3)}{8\Delta_2(1+2\Delta_2)},\\
        c_{112}&=(\Delta_t-2\Delta_2-1)\frac{c_{02}(\Delta_t-2\Delta_3)-4c_{62}\Delta_2}{4\Delta_2(1+2\Delta_2)}.
    \end{split}
\end{align}
Notice that in contrast with the $\mathcal{N}=1$ three point function \eqref{Neq13pointSol} which had two free parameters, its $\mathcal{N}=2$ counterpart has four ($c_{01},c_{02},c_{61},c_{62}$). This motivates us to check if there exists a double copy like we found for the two point case \eqref{DoubleCopy}. However, when we explicitly checked, we found no such double copy relation. Perhaps we need to employ variables similar to the Grassmann twistor variables of \cite{Jain:2023idr} for a double copy relation to be manifest. We leave such an exercise as well as the analysis of higher point functions and higher supersymmetry to the future.
\section{Discussion and Future Directions}\label{sec:Discussion}
In this paper, we have found a general form for \textit{n}-point correlation functions in momentum space conformal quantum mechanics. These take the form of the Lauricella hypergeometric function with $n-3$ undetermined parameters (see \eqref{nPointSol}), which exactly coincides with the number of cross ratios in real space. Such a representation has not yet been achieved in the literature except in regimes with special kinematics or extra symmetries \cite{Coriano:2019nkw} in generic dimensions. Our result in $d=1$, however, relies only on conformal invariance and requires no additional assumptions. As mentioned in the introduction, the CQM conformal Ward identities coincide with the (anti-)holomorphic Ward identities in two-dimensional CFT \cite{Belavin:1984vu}. Thus, by taking a sum of the products of our solutions (with one solution being interpreted as holomorphic and the other as anti-holomorphic), we obtain candidate two-dimensional CFT momentum space correlators. It would thus be interesting to put our results for correlators as well the momentum space conformal partial waves \eqref{CPW2} to the test in that arena. It would also be interesting to see what extra constraints are obtained by demanding the Virasoro symmetry. \\

Another avenue to pursue is the holographic computation of CFT$_1$ correlators, i.e, in (A)dS$_2$. It would be interesting to see how the requirement of holography and the consistency (unitarity, locality, absence of un-physical singularities, etc...) \cite{Baumann:2022jpr} of the bulk theory further constrains our results. This would be especially useful, for instance in the context of the SYK model and it's holographic bulk dual \cite{Gross:2017aos,Rosenhaus:2018dtp}. Determining the flat space limit of our correlators by taking the limit of large scaling dimensions and the (A)dS radius to infinity (see for instance, \cite{Li:2021snj}) is also another avenue of interest. The results can be matched with two-dimensional scattering amplitudes. The applicability of our results to celestial \cite{Pasterski:2021raf} and carrollian \cite{Bagchi:2022emh} flat space holography presents another important frontier. For instance, interesting connections between higher dimensional Carrollian and $0+1$ dimensional quantum mechanical models have been found in \cite{Kasikci:2023zdn} and would be interesting to explore further.\\

Another point that we have attempted to clarify is the existence of multiple solutions to the momentum space conformal Ward identities. There have been given a few interpretations in the literature \cite{Gillioz:2021sce,Jain:2022uja}, and in this paper, we provide a natural explanation for the same: The different solutions correspond to the Fourier transform of the different possible time orderings.\\

Finally, it would also be interesting to see if we can make similar statements in higher dimensions, i.e, to check if the conformal Ward identities can be mapped to some known system of PDEs.
\subsection*{Acknowledgments}
We would like to thank Fernando Alday, M. Ali, Sachin Jain, Sunil Mukhi, Kostas Skenderis and Farman Ullah for valuable discussions. D K.S would also like to thank the organizers of the 18th Kavli Asian Winter School (YITP-W-23-13) for providing a stimulating environment and Precision Wires India Ltd for String Theory and Quantum Gravity research at IISER Pune for making the trip possible.  S.Y acknowledges support from INSPIRE fellowship from the Department of Science and Technology, Government of India. 
\appendix
\section{The Super Conformal Algebra and Ward Identities}
\subsection{$\mathcal{N}=1$ Superconformal Algebra and Ward Identities}\label{appendix:Neq1}
The generators of the $\mathcal{N}=1$ superconformal algebra obey the following (anti)commutation relations: 
\begin{align}
    [D,H]=-iH,~~~[D,K]&=iK,~~~[K,H]=-2iD,\notag\\
    \{Q,Q\}=H,~~~\{S,S\}&=-K,~~~\{Q,S\}=iD,\notag\\
    [D,Q]=-\frac{i}{2}Q,~~~[D,S]=\frac{i}{2}S,~&~[K,Q]=-S,~~~[H,S]=-Q.
\end{align}
Their action on primary operators is as follows:
\begin{align}\label{Neq1Susygenerators}
\begin{split}[H,\mathbf{O}_{\Delta}]&=\omega \mathbf{O}_{\Delta},\\
    [D,\mathbf{O}_{\Delta}]&=-i\left(\omega\frac{\partial}{\partial \omega}+(1-\Delta)-\frac{1}{2}\theta \frac{\partial}{\partial \theta}\right)\mathbf{O}_{\Delta},\\
    [K,\mathbf{O}_{\Delta}]&=-\left(\omega \frac{\partial^2}{\partial\omega^2}+2(1-\Delta)\frac{\partial}{\partial \omega}-\theta\frac{\partial}{\partial \theta}\frac{\partial}{\partial \omega}\right)\mathbf{O}_{\Delta},\\
    [Q,\mathbf{O}_{\Delta}]&=\left(\frac{\partial}{\partial \theta}+\frac{\theta}{2}\omega\right)\mathbf{O}_{\Delta} ,\\
[S,\mathbf{O}_{\Delta}]&=\left(\frac{\partial}{\partial \theta}\frac{\partial}{\partial \omega}+\left(\frac{1}{2}-\Delta\right)\theta+\frac{\theta \omega}{2}\frac{\partial}{\partial \omega}\right)\mathbf{O}_{\Delta}.
\end{split}
\end{align}
\subsection{$\mathcal{N}=2$ Superconformal Algebra and Ward Identities}\label{appendix:Neq2}
The super lie algebra obeyed by the $\mathcal{N}=2$ superconformal generators can be found, for instance, in \cite{Okazaki:2015pfa}.
The action of the $\mathcal{N}=2$ superconformal algebra generators on primary operators is as follows:
\begin{align}[H,\mathbf{O}_{\Delta}]&=\omega \,\mathbf{O}_{\Delta}\label{Ne2susyhaction},\\
    [D,\mathbf{O}_{\Delta}]&=-i\left(\omega\frac{\partial}{\partial \omega}+(1-\Delta)-\frac{1}{2}\theta \frac{\partial}{\partial \theta}-\frac{1}{2}\bar\theta \frac{\partial}{\partial \bar\theta}\right)\mathbf{O}_{\Delta}\label{Ne2susydaction},\\
    [K,\mathbf{O}_{\Delta}]&=-\left(\omega \frac{\partial^2}{\partial\omega^2}+2(1-\Delta)\frac{\partial}{\partial \omega}-\theta\frac{\partial}{\partial \theta}\frac{\partial}{\partial \omega}-\bar\theta\frac{\partial}{\partial\bar \theta}\frac{\partial}{\partial \omega}\right)\mathbf{O}_{\Delta},\label{Ne2susykaction}\\
    [Q,\mathbf{O}_{\Delta}]&=\left(\frac{\partial}{\partial \theta}+\frac{\Bar{\theta}}{2}\omega\right)\mathbf{O}_{\Delta},\quad[\Bar{Q},\mathbf{O}_{\Delta}]=\left(\frac{\partial}{\partial\Bar{ \theta}}+\frac{\theta}{2}\omega\right)\mathbf{O}_{\Delta} \label{Ne2susyqaction},\\
[S,\mathbf{O}_{\Delta}]&=\left(\frac{\partial}{\partial \theta}\frac{\partial}{\partial \omega}+\left(\frac{1}{2}-\Delta\right)\Bar{\theta}+\frac{\Bar{\theta} \omega}{2}\frac{\partial}{\partial \omega}-\frac{\Bar{\theta}\theta}{2}\frac{\partial}{\partial\theta}\right)\mathbf{O}_{\Delta},\label{Neq2susysaction}\\ [\Bar{S},\mathbf{O}_{\Delta}]
&=\left(\frac{\partial}{\partial \Bar{\theta}}\frac{\partial}{\partial \omega}+\left(\frac{1}{2}-\Delta\right)\theta+\frac{\theta \omega}{2}\frac{\partial}{\partial \omega}-\frac{\theta\Bar{\theta}}{2}\frac{\partial}{\partial\Bar{\theta}}\right)\mathbf{O}_{\Delta}\label{Ne2susysbaraction},\\
[R,\mathbf{O}_{\Delta}]&=\left(\theta\frac{\partial}{\partial\theta}-\Bar{\theta}\frac{\partial}{\partial\Bar{\theta}}\right)\mathbf{O}_{\Delta}.\label{Ne2susyRaction}
\end{align}
These imply the following Ward identities for the correlation functions:
\begin{align}\label{WardIdNeq2}
\begin{split}
\sum_{i=1}^n \omega_i \,f_n(\omega_1,\theta_1,\bar\theta_1;\dots;\omega_n,\theta_n,\bar\theta_n)=0,\\
 \sum_{i=1}^n\left(\omega_i\frac{\partial}{\partial \omega_i}+(1-\Delta_i)-\frac{1}{2}\theta_i \frac{\partial}{\partial \theta_i}-\frac{1}{2}\bar\theta_i \frac{\partial}{\partial \bar\theta_i}\right)\,f_n(\omega_1,\theta_1,\bar\theta_1;\dots;\omega_n,\theta_n,\bar\theta_n)=0,\\
    \sum_{i=1}^n\left(\omega_i \frac{\partial^2}{\partial\omega_i^2}+2(1-\Delta_i)\frac{\partial}{\partial \omega_i}-\theta_i\frac{\partial}{\partial \theta_i}\frac{\partial}{\partial \omega_i}-\bar\theta_i\frac{\partial}{\partial \bar\theta_i}\frac{\partial}{\partial \omega_i}\right)\,f_n(\omega_1,\theta_1,\bar\theta_1;\dots;\omega_n,\theta_n,\bar\theta_n)=0,\\
    \sum_{i=1}^n\left(\frac{\partial}{\partial \theta_i}+\frac{\Bar{\theta}_i}{2}\omega_i\right)\,f_n(\omega_1,\theta_1,\bar\theta_1;\dots;\omega_n,\theta_n,\bar\theta_n)=0,\\
    \sum_{i=1}^n\left(\frac{\partial}{\partial\Bar{ \theta}_i}+\frac{\theta_i}{2}\omega_i\right)\,f_n(\omega_1,\theta_1,\bar\theta_1;\dots;\omega_n,\theta_n,\bar\theta_n)=0 ,\\
\sum_{i=1}^n\left(\frac{\partial}{\partial \theta_i}\frac{\partial}{\partial \omega_i}+\left(\frac{1}{2}-\Delta_i\right)\Bar{\theta}_i+\frac{\Bar{\theta}_i \omega_i}{2}\frac{\partial}{\partial \omega_i}-\frac{\Bar{\theta}_i\theta_i}{2}\frac{\partial}{\partial\theta_i}\right)\,f_n(\omega_1,\theta_1,\bar\theta_1;\dots;\omega_n,\theta_n,\bar\theta_n)=0,\\ \sum_{i=1}^n\left(\frac{\partial}{\partial \Bar{\theta}_i}\frac{\partial}{\partial \omega_i}+\left(\frac{1}{2}-\Delta_i\right)\theta_i+\frac{\theta_i \omega_i}{2}\frac{\partial}{\partial \omega_i}-\frac{\theta_i\Bar{\theta}_i}{2}\frac{\partial}{\partial\Bar{\theta}_i}\right)\,f_n(\omega_1,\theta_1,\bar\theta_1;\dots;\omega_n,\theta_n,\bar\theta_n)=0,\\
\sum_{i=1}^n\left(\theta_i\frac{\partial}{\partial\theta_i}-\Bar{\theta}_i\frac{\partial}{\partial\Bar{\theta}_i}\right)\,f_n(\omega_1,\theta_1,\bar\theta_1;\dots;\omega_n,\theta_n,\bar\theta_n)=0.
\end{split}
\end{align}
where, $f_n(\omega_1,\theta_1,\bar\theta_1;\dots;\omega_n,\theta_n,\bar\theta_n)$ is a general $\mathcal{N}=2$, $n$-point function.
\section{Series expansions for the Lauricella functions}\label{appendix:SeriesFormulae}
A series expansion for the general Lauricella function of $m$ variables is the following \cite{matsumoto}:
\begin{align}
    &E_A^{(m)}(a,b_1,\cdots,b_m,c_1,\cdots,c_m;-x_1,\cdots,-x_m)=\sum_{m_i\in\mathbb{N}_0}\frac{(a)_{m_1+\cdots+m_n}\prod_{i=1}^{m}(b_i)_{m_i}}{\prod_{i=1}^{m}(c_i)_{m_i}\prod_{i=1}^{m}m_i!}\prod_{i=1}^{m}x_i^{m_i}.
\end{align}
$\mathbb{N}_0=\{0,1,2,\cdots\}$ and $\displaystyle(a)_n=\frac{\Gamma(a+n)}{\Gamma(a)}$ is the rising Pochammer symbol. The above series converges when $\displaystyle \sum_{i=1}^{m}|x_i|<1$. As a special case we obtain $~_2 F_1$, when $m=1$; Appell $F_2$, when $m=2$; $E_A^{(3)}$, when $m=3$ and so on.
\section{Useful identities involving hypergeometric functions}\label{appendix:hypergeometric}
Some useful identities involving the hypergeometric functions that we used in the main text are:
 \begin{align}
     \frac{2\Delta_2-1}{\Delta_t-1} ~_2F_1(1-\Delta_t,1-\Delta_t+2\Delta_1,1-2\Delta_2;-x)&+x\, ~_2F_1(2-\Delta_t,1-\Delta_t+2\Delta_1,2-2\Delta_2;-x)\notag\\=\frac{2\Delta_2-1}{\Delta_t-1}& ~_2F_1(1-\Delta_t,2\Delta_1-\Delta_t,1-2\Delta_2;-x).
     \end{align}
     \begin{align}
   \frac{2\Delta_2}{2\Delta_1+2\Delta_2-\Delta_t}~_2F_1(&1+2\Delta_2-\Delta_t,2\Delta_1+2\Delta_2-\Delta_t,2\Delta_2;-x)\notag\\
   &-~_2F_1(1+2\Delta_2-\Delta_t,1+2\Delta_1+2\Delta_2-\Delta_t,1+2\Delta_2;-x)\notag\\&=\frac{\Delta_t-2\Delta_1}{2\Delta_1+2\Delta_2-\Delta_t}~_2F_1(1+2\Delta_2-\Delta_t,2\Delta_1+\Delta_2-\Delta_t,1+2\Delta_2;-x).
 \end{align}

\bibliographystyle{JHEP}
\bibliography{bibliography}

\end{document}